\RequirePackage{rotating}
\documentclass[useAMS,usenatbib]{mn2e}

\usepackage{lscape}

\usepackage{amssymb}
\usepackage{graphicx}
\usepackage{rotating, caption}
\usepackage{epstopdf}
\usepackage{times}	
\usepackage{float}   
\usepackage{comment} 
\usepackage{amsmath} 

\title[Mon R2 $Herschel-SPIRE$ survey]{A $Herschel-SPIRE$ Survey of the Mon~R2 Giant Molecular Cloud: Analysis of the Gas Column Density Probability Density Function\thanks{{\it Herschel} is an ESA space observatory with science instruments provided by European-led Principal Investigator consortia and with important participation from NASA.}}\author[Pokhrel et al.]
{R. Pokhrel$^{1}$\thanks{E-MAIL: rpokhrel@astro.umass.edu}, 
R. Gutermuth$^{1}$, 
B. Ali$^{2}$, 
T. Megeath$^{3}$, 
J. Pipher$^{4}$, 
P. Myers$^{5}$, 
W. J. Fischer$^{6}$
\thanks{NASA Postdoctoral Program Fellow},
\newauthor 
T. Henning$^{7}$, 
S. J. Wolk$^{5}$, 
L. Allen$^{8}$, 
J. J. Tobin$^{9}$
\\
$^{1}$University of Massachusetts, Amherst, MA 01003, USA\\
$^{2}$Space Science Institute, Boulder, CO 80301, USA\\
$^{3}$University of Toledo, Toledo, OH 43606, USA\\
$^{4}$University of Rochester, Rochester, NY 14627, USA\\
$^{5}$CFA- Harvard University, Cambridge, MA 02138, USA\\
$^{6}$NASA's Goddard Space Flight Center, Greenbelt, MD 20771, USA\\
$^{7}$MPIA- Heidelberg, K\"{o}nigstuhl 17, 69117 Heidelberg, Germany\\
$^{8}$NOAO, Tucson, AZ 85719, USA\\
$^{9}$P.O. Box 9513, NL-2300 RA Leiden, The Netherlands\\
}

\begin{document}

\date{Accepted YEAR MONTH DAY. Received YEAR MONTH DAY; in original form YEAR MONTH DAY}

\pagerange{\pageref{firstpage}--\pageref{lastpage}} \pubyear{2015}

\maketitle

\label{firstpage}

\begin{abstract}
We present a far-IR survey of the entire Mon R2 GMC with $Herschel-SPIRE$ cross-calibrated with $Planck-HFI$ data. We fit the SEDs of each pixel with a greybody function and an optimal beta value of 1.8. We find that mid-range column densities obtained from far-IR dust emission and near-IR extinction are consistent. For the entire GMC, we find that the column density histogram, or N-PDF, is lognormal below $\sim$10$^{21}$ cm$^{-2}$. Above this value, the distribution takes a power law form with an index of -2.16. We analyze the gas geometry, N-PDF shape, and YSO content of a selection of subregions in the cloud. We find no regions with pure lognormal N-PDFs. The regions with a combination of lognormal and one power law N-PDF have a YSO cluster and a corresponding centrally concentrated gas clump. The regions with a combination of lognormal and two power law N-PDF have significant numbers of typically younger YSOs but no prominent YSO cluster. These regions are composed of an aggregate of closely spaced gas filaments with no concentrated dense gas clump. We find that for our fixed scale regions, the YSO count roughly correlates with the N-PDF power law index. The correlation appears steeper for single power law regions relative to two power law regions with a high column density cut-off, as a greater dense gas mass fraction is achieved in the former. A stronger correlation is found between embedded YSO count and the dense gas mass among our regions.

\end{abstract}

\begin{keywords}
ISM: individual objects (Mon~R2) -- ISM: clouds -- ISM: structure
\end{keywords}

\section{Introduction} \label{sec:intro}

Stars form within the dense regions of diffuse molecular clouds, but the physical processes that determine the locations, rate, and efficiency of star formation are poorly understood. Based on preliminary results of the recent $Herschel$ Gould Belt survey (HGBS), \cite{2010A&A...518L.102A} summarized the current picture of structure formation in clouds as a two step process: first, a network of dense filaments are formed due to large-scale magneto-hydrodynamic turbulence, and then fragmentation occurs as gravity wins over turbulence thereby forming prestellar cores. The structure of the dense gas is affected by motions induced by supersonic turbulence \citep{1997MNRAS.288..145P}, self-gravity of gas \citep{2000ApJ...535..887K} and magnetic fields \citep{2012MNRAS.423.2680M} inside the cloud. However, the role of each physical process in structure formation is still debated \citep{2007ARA&A..45..565M}.

Recent work suggests that aggregate column density diagnostics, such as the column density probability distribution function (N-PDF), which gives the probability of a region to have a column density within [$N$, $N$ + d$N$], are key to the identification of structure formation caused by different dominant physics (e.g., \citealt{2009A&A...508L..35K, 2010MNRAS.405L..56B}). The cloud regions dominated by different physical phenomena exhibit N-PDFs of differing functional form. Aggregate N-PDFs of quiescent clouds with very little star formation activity display a lognormal form, suggestive of structures formed by supersonic turbulence \citep{2010MNRAS.405L..56B}. Active star forming regions exhibit some evidence of lognormal distribution at low densities as well, but also tend to have a power law excess from medium to high column densities, where gas self gravity may be the dominant physics driving the cloud structure formation \citep{2009A&A...508L..35K}. This mixed picture has been brought forward both by observations (e.g., \citealt{2011A&A...533A..94H}) and theory (e.g., \citealt{2014prpl.conf...77P}), but only recently have datasets reached the degree of quality necessary to allow discriminating tests of this picture.

Among a few methods to map the distribution of column densities are mm-line emission features, extinction of background stars by dust, and thermal dust emission (c.f., \citealt{2005ApJ...634..442S}). There are advantages and pitfalls to each process \citep{2009ApJ...692...91G}. For using line emission features, molecular tracers are mostly constrained by the optical depth effect of clouds, chemistry, and abundance variations more broadly. On the other hand, the dust extinction suffers from selection effects as it depends on the detection of background stars. Thus, the lack of good photometry of weak stellar sources in high extinction regions can make this process biased towards lower density regions. In contrast, dust emission is free of these biases and allows us to make column density maps with a very wide dynamic range. The advent of $Herschel$, with its unprecedented angular resolution and sensitivity in the far-IR, has enabled dust emission mapping of substantially greater quality than has been previously available over large areas of sky \citep{2010A&A...518L.102A}.

Dust emission in the ISM is best modelled by an ensemble of emitting particles that spans a substantial range of sizes, compositions, and temperatures \citep{1978BAAS...10Q.463D}. For each line of sight, there can be different emissivity properties for these emitting particles (dust grains) with different properties. Depending upon the availability of data, different methods can be used to estimate emissivity for each line of sight \citep{2013ApJ...767..126S}. Recent efforts to estimate column density, temperature, and dust emissivity generally adopt a modified blackbody (greybody) model with wavelength dependent emissivity to represent the aggregate dust emission along each line of sight (e.g., \citealt{1994ApJS...95..457W}).

In this work, we present an analysis of a new survey of the Mon~R2 giant molecular cloud (GMC) with $Herschel$- $SPIRE$.  Our goal is to characterize the column density and temperature structure of the cloud, which is more distant, physically larger than the clouds in Gould's Belt, and more actively forming stars. In $\S$.\ref{sec:intro}, we introduce the research with a literature review of Mon~R2 GMC. In $\S$.\ref{dataanalysis}, we explain our observations and data reduction procedure along with an overview on the technique we implemented in understanding the dust properties of Mon~R2. We analyse the column density distribution in $\S$.\ref{coldendist}. Finally, the conclusion of the paper is summarized in $\S$.\ref{conclusion}.

\subsection{Mon~R2 Giant Molecular Cloud} \label{monr2}

The Mon~R2 region was originally identified as a group of reflection nebulae in the constellation of Monoceros. \cite{1920ApJ....52....8S} initially identified stars that may be exciting the nebulae and are responsible for the extended emission of the cloud. The first detailed spectroscopic and photometric study of Mon~R2 nebulae was done by \cite{1968AJ.....73..233R} who discovered that the illuminating associated stars are mainly B-type stars. \cite{1968AJ.....73..233R} estimated the distance to the cloud as 830$\pm$50 pc which was later re-confirmed by \cite{1976AJ.....81..840H} by fitting the zero-age main sequence locus from \cite{1963bad..book..204J} to dereddened $UBV$ photometry.

\cite{1974ApJ...194L.103L} reported the first $^{12}$CO (J=1-0) detection in Mon~R2 and \cite{1975ApJ...199...79K} showed that at least five of the reflection nebulae are associated with local maxima in $^{12}$CO maps. The cloud was first mapped in its entirety in $^{12}$CO by \cite{1986ApJ...303..375M} who surveyed $\sim$ 3$^{\circ}$ $\times$ 6$^{\circ}$ (44 pc $\times$ 55 pc) region of the GMC. Peaks in molecular emission corresponding to the location of reflection nebulae are traced by $^{13}$CO \citep{1999ApJ...524..895M}. \cite{1992PhDT.........2X} estimated the mass 4 $\times$ 10$^4$ M$_{\sun}$ for the cloud using $^{12}$CO.

\cite{2000AJ....120.3139C} used the $2MASS$ point source catalog to identify compact stellar clusters, finding four clusters based on enhancements in stellar surface density relative to the field star population. These four clusters are associated with the Mon~R2 core, GGD 12-15, GGD 17, and IRAS 06046-0603 as shown in figure \ref{fig:pxw}. More recent works include the analysis of stellar distributions by K-band number counts and structure analysis, using near-IR data obtained with FLAMINGOS on the MMT and including SCUBA 850 $\micron$ data \citep{2005ApJ...632..397G}. \cite{2007AJ....134.2020H} used the Wide-Field Camera on UKIRT in the 2.12 $\micron$ filter centered on the H$_2$ 1-0 $S(1)$ emission line and discovered 15 H$_2$ jets in Mon~R2, confirming most of the discoveries using archival $Spitzer$-$IRAC$ 4.5 and 8.0 $\micron$. This work further asserted that the jets may be associated with an episode of star formation in Mon~R2 triggered by the large central outflow. Further analysis by \cite{2011ApJ...739...84G} reports a power law correlation with a slope 2.67 for the Mon~R2 cloud between the local surface densities of $Spitzer$ identified YSOs ($\sim$ 1000) and the column density of gas as traced by near-IR extinction.

\section[]{Data Reduction and Analysis} \label{dataanalysis}

We surveyed the entire Mon~R2 GMC with parallel scan-map mode with the ESA $Herschel$ Space Observatory \citep{2010A&A...518L...1P} using both the Photoconductor Array Camera and Spectrometer, $PACS$, \citep{2010A&A...518L...2P} and the Spectral and Photometric Imaging REceiver, $SPIRE$, \citep{2010A&A...518L...3G}. The target name as designated in the $Herschel$ Science Archive (HSA) is Mon~R2-3 with corresponding OBSIDs 1342267715 and 1342267746. For our analysis, we used only $SPIRE$ data as we could not recover large scale structure in the $PACS$ 160 $\micron$ map reliably because it was found to be harshly contaminated.  At large scales, $PACS$ 70 $\micron$ can not be used for studying extended emission. $Herschel$ observed Mon~R2 on 15 March 2013 covering the area of 4.30$^{\circ}$ by 4.36$^{\circ}$ and centered on 06h 08m 46.90s $\textit{RA}$(J2000), - 06$^{\circ}$ 23$'$ 12.33$''$ $\textit{Dec}$(J2000) with position angle of 268.09$^{\circ}$. We obtained level-2 $SPIRE$ data products at 250 $\micron$, 350 $\micron$ $\&$ 500 $\micron$ in both in-scan and cross-scan mode, i.e., in orthogonal scan directions to help with mitigation of scanning artifacts, at the scanning speed of 60$''$ s$^{-1}$. We reduced $SPIRE$ observations using $Herschel$ Interactive Processing Environment $(HIPE)$ \citep{2010ASPC..434..139O}, version 11.0.1. We adjusted standard pipeline scripts to construct combined maps recovering the extended emission from the two sets of scans. 

\begin{figure}
\centering
\includegraphics[width=3.1in]{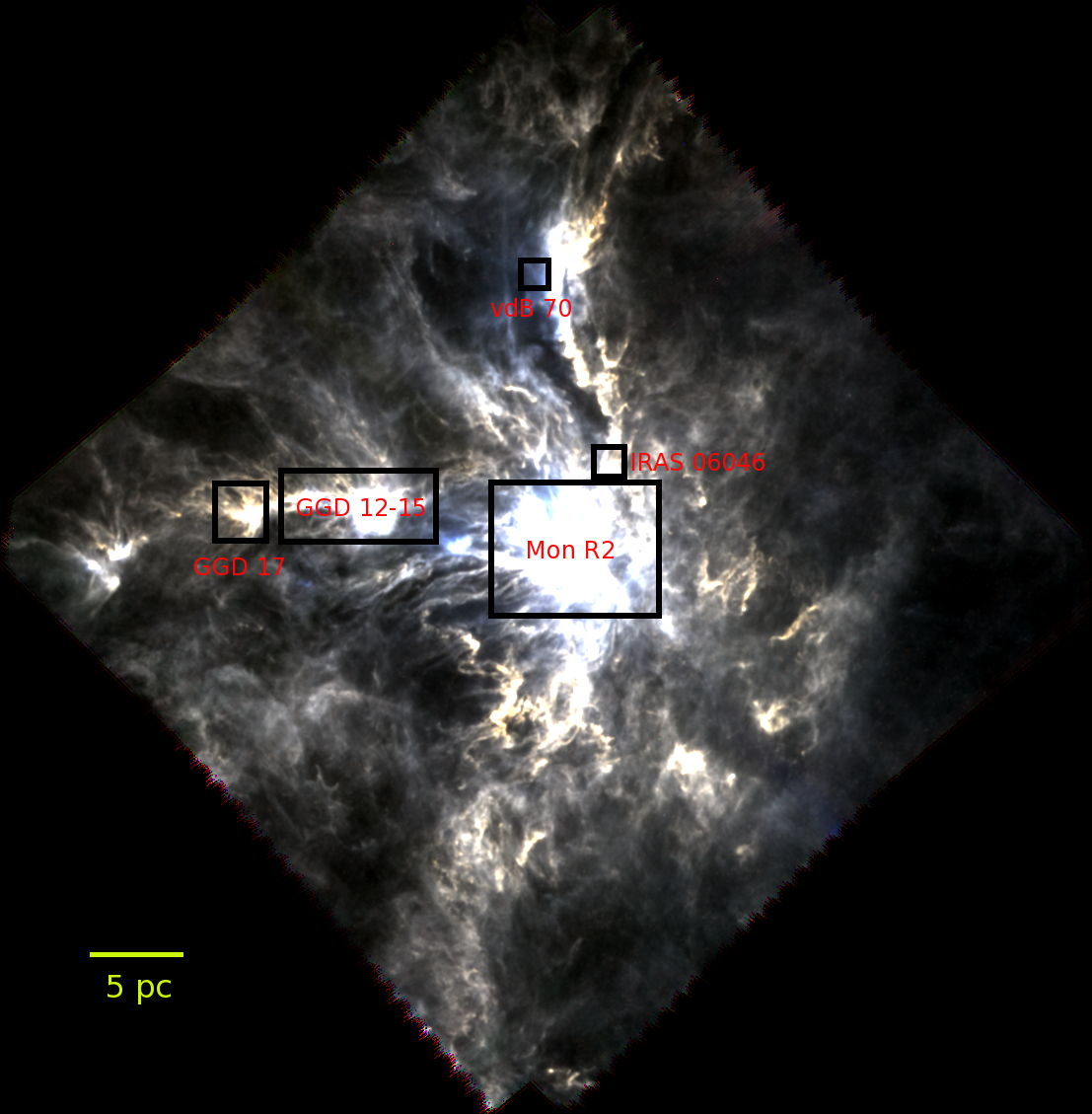}
\caption{False colour RGB image of the final set of $SPIRE$ images where $B$ = 250 $\micron$, $G$ = 350 $\micron$ $\&$ $R$ = 500 $\micron$ respectively, after calibrating with $Planck$-$HFI$.
\label{fig:pxw}}
\end{figure}

We used $Planck$ High Frequency Instrument ($HFI$) \citep{2011A&A...536A...4P} maps to obtain an absolute calibration for the $SPIRE$ maps. $Planck$-$HFI$ is a bolometric detector array designed to produce high sensitivity measurements covering the full sky in the wavelength range 0.3 to 3.6~mm. $Herschel$-$SPIRE$ detectors are only sensitive to relative variations and absolute brightness can not be known but $Planck$-$HFI$ sets an absolute offset to its maps. Both $SPIRE$ and $HFI$ share two channels with overlapping wavebands.  This is an advantage in calibrating $SPIRE$ maps (c.f., \citealt{2010A&A...518L..88B}). We used $Planck$ $HFI$-545 and $HFI$-857 for this purpose, each with an angular resolution of 5$'$. We used the standard $HIPE$ method to calibrate $Herschel$ maps. For this, $HIPE$ convolves the $SPIRE$ maps with the $PLANCK$ beam in the area of interest and sets the median $SPIRE$ image flux level to be equal to the median $PLANCK$ level in the same area. All three $SPIRE$ maps are absolute calibrated using the $HFI$ emission to greybody conversion and colour correction for $SPIRE$ assuming a greybody source spectrum, $I_S \sim \nu^{\beta}B_{\nu}(T)$ (see $SPIRE$ handbook\footnote{$http://herschel.esac.esa.int/Docs/SPIRE/html/spire\_om.html$} for details). For the $SPIRE$ reduction, we applied relative gains, ran the de-striper in each band and then applied the zero-point correction using the standard $HIPE$ technique.  Then we combined the final data for each bandpass into three mosaics used in the analysis discussed below. The RGB image of these three wavebands are shown in figure \ref{fig:pxw}.

The three bands in $SPIRE$ have angular resolutions of 18$''$, 25$''$ and 36$''$. In order to fit a modified blackbody function to the data from these images, we convolved the higher resolution images with a 2D Gaussian kernel of an appropriate full width at half maximum (FWHM) to the resolution of the 500 $\micron$ data using the recipe from \cite{2011PASP..123.1218A}. It provides appropriate kernels for most of the space and ground based telescopes for the purpose of matching resolution in two sets of images. The convolved images were then regridded to 14$''$ pixel size corresponding to $SPIRE$ 500 $\micron$ using the {\it hastrom} routine from the Interactive Data Language (IDL) Astronomy Users Library \citep{1993ASPC...52..246L} so that a given pixel position in each image corresponds to the same position on the sky.

\begin{figure}
\centering
\includegraphics[width=3.2in]{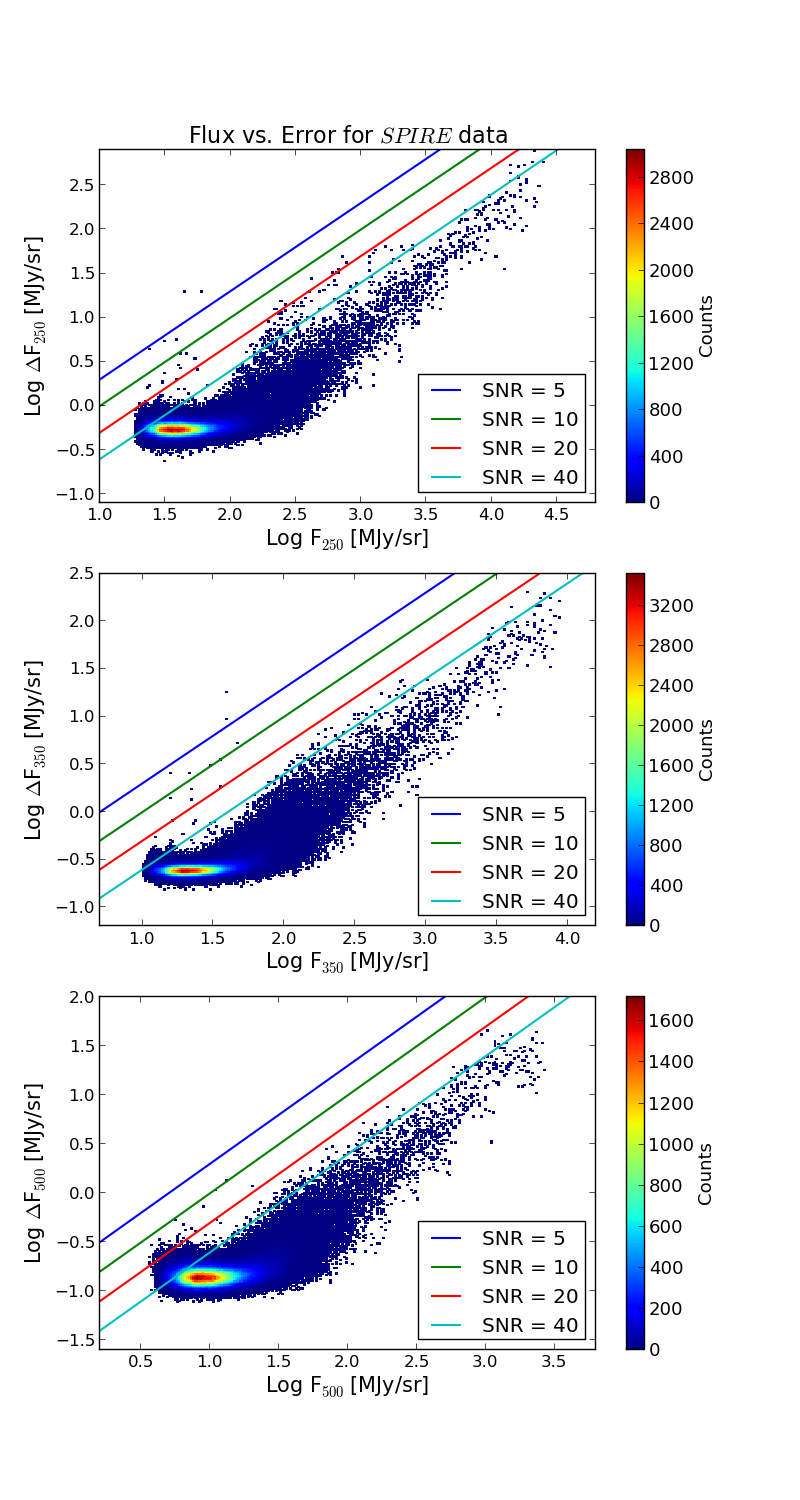}
\caption{Flux vs. Error plot for $SPIRE$ wavebands showing the actual flux values with their 1-$\sigma$ uncertainties, presented in the form of a 2D histogram. Synthetic lines for different SNR values are overplotted for each waveband.
\label{fig:fluxvserr}}
\end{figure}

In figure \ref{fig:fluxvserr}, we present the resulting flux vs. uncertainty plots for all $SPIRE$ bands as 2D histograms, with fiducial signal to noise ratio (SNR) lines overplotted. The vast majority of our data have high SNR.  Our basis for the selection of high quality data points for subsequent analysis includes a requirement of SNR $>$ 10 in all three bands and good observing coverage based on the HSA-provided coverage maps.

\subsection{Estimating dust properties} \label{dust}

Cold dust emission in nearby molecular clouds is a thermal process and is generally modelled with a blackbody spectrum modified by a frequency-dependent emissivity (e.g., \citealt{1983QJRAS..24..267H}). In general, the radiative transfer equation governing the emission $I_{\nu}$ for a modified blackbody spectrum is:

\begin{equation} \label{eq1}
I_{\nu} = B_{\nu}(T) \times (1 - e^{-\tau_{\nu}}) + I_{\nu}^{\rm{back}}e^{-\tau_{\nu}} + I_{\nu}^{\rm{fore}}
\end{equation}

where $B_{\nu}(T)$ is the Planck function for a perfect blackbody of temperature T in Kelvin and $\tau_{\nu}$ is the opacity of the cloud at corresponding frequency $\nu$. Equation \ref{eq1} for the optically thin case takes the form:  

\begin{equation} \label{eq1s}
I_{\nu} = B_{\nu}(T) \times \tau_{\nu} + I_{\nu}^{\rm{com}}
\end{equation}

where $I_{\nu}^{\rm{com}}$ is the combined foreground and background emission which can be neglected in our case due to the position of the cloud, several degrees away from the Galactic plane. Since, $\tau_{\nu}$ = $\kappa_{\nu}$ $\Sigma$; $\Sigma$ being the mass surface density and $\kappa_{\nu} \propto \nu^{\beta}$ where $\beta$ is the dust emissivity power law index, equation \ref{eq1s} takes the following form:

\begin{equation} \label{eq2}
I_{\nu} = \kappa_{\nu_0}(\nu/\nu_0)^{\beta}B_{\nu}(T)\Sigma
\end{equation}

where $\kappa_{\nu_0}$ is a reference dust opacity per unit gas and dust mass at a reference frequency $\nu_0$. We took $\kappa_{\nu_0}$ = 2.90 cm$^2$/gm for $\nu_0$ corresponding to the longest observed wavelength, 500 $\micron$, following the OH-4 model (\citet{1977ApJ...217..425M} distribution for dust grains in the ISM: $f(a) \propto a^{-3.5}$, with thin ice mantles on dust grains and no coagulation) from \citet{1994A&A...291..943O}. $T$ is the dust temperature and $\Sigma$ = $\mu m_{H}N(H_2)$ where $\mu$ is the mean molecular weight per unit hydrogen mass $\sim$ 2.8 for a cloud with 71\% molecular hydrogen, 27\% helium and 2\% metals \citep{2013ApJ...767..126S}, $m_H$ is the mass of single hydrogen atom and $N(H_2)$ is the gas column density. Hence, we are observing the dust emission and using it as a probe to compute the gas column density. We assume the canonical gas-to-dust ratio of 100 \citep{1995A&A...293..889P} for our purpose.

\subsubsection{Dust spectral index, $\beta$} \label{dustbeta}

Our goal is to fit the spectral energy distribution (SED) of individual pixels with equation \ref{eq2}. Since we could not rely on the $PACS$ data, it limited the available data to three $SPIRE$ wavebands only. This led to the problem of fitting the observations with three data points with an equation of three unknowns. \cite{2010A&A...518L.102A}, \cite{2012A&A...540A..10S} and \cite{2011A&A...529L...6A} assume $\beta$ to be a constant between 1.5 (hotter regions) and 2 (colder regions). We used the $SPIRE$ flux ratios to estimate the most appropriate value of $\beta$ so that it best represents the whole cloud complex.

\begin{figure}
\centering
\includegraphics[width=3.5in]{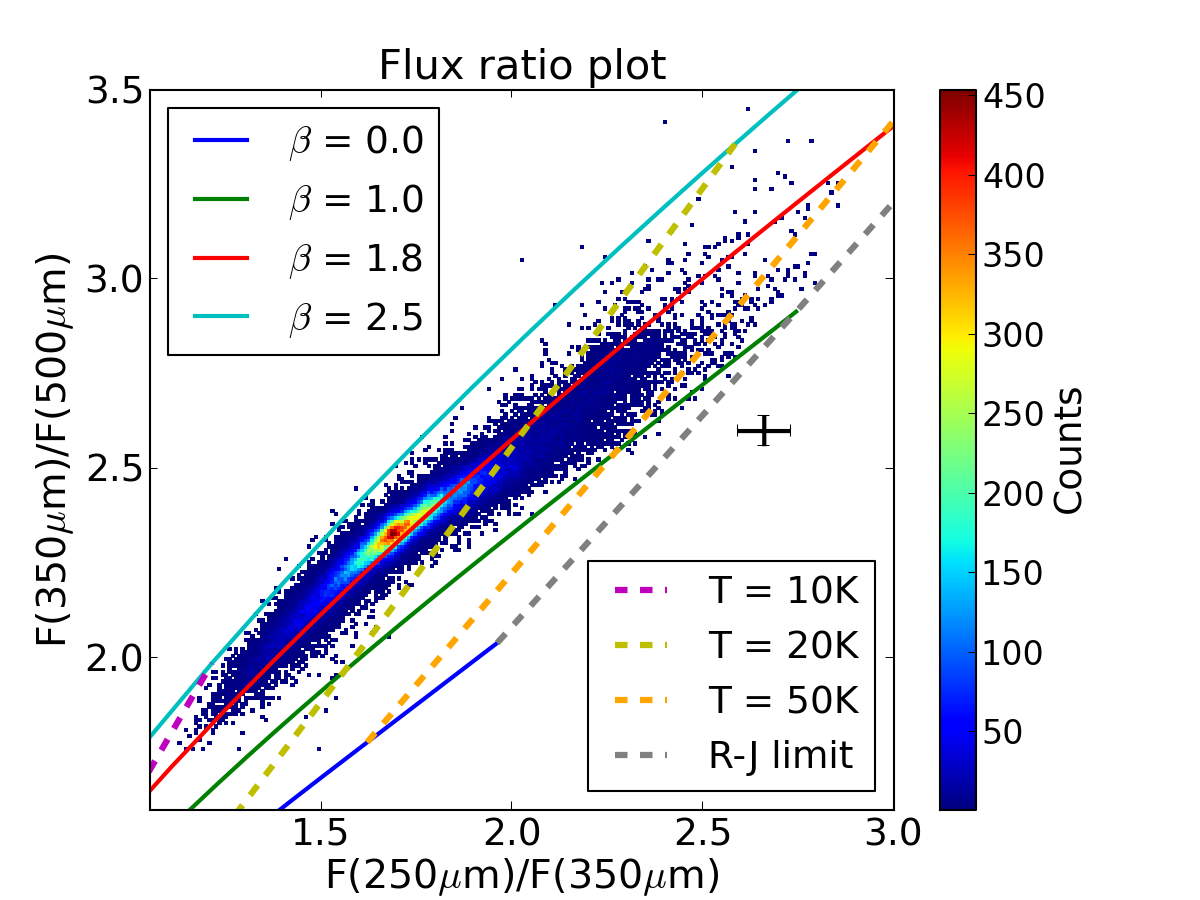}
\caption{$F_{250}/F_{350}$ Vs. $F_{350}/F_{500}$ plot as a 2D histogram, overlaid with theoretical greybody plots for different $\beta$ and temperature ranges, showing unique values for both $\beta$ and temperature for each pixel. The data are best matched across the entire space with $\beta$ = 1.8. The black cross in the plot represents the typical error, which is the median error of the overall distribution of flux ratio points.
\label{fig:fluxRatio}}
\end{figure}

In Figure \ref{fig:fluxRatio} we plot the ratio of fluxes between 250 $\micron$ and 350 $\micron$ on the x-axis and the ratio of fluxes between 350 $\micron$ and 500 $\micron$ on the y-axis. Equation \ref{eq2} is used to compute the reference flux ratios at different wavebands for a given choice of $\beta$ and temperature. To our benefit, the column density and emissivity calibration constant cancel in this ratio-space. For frequencies $\nu_1$ and $\nu_2$, the equation reduces to a simple form:

\begin{equation} \label{eq4}
\frac{I_{\nu_1}}{I_{\nu_2}} = \Bigg (\frac{\nu_1}{\nu_2} \Bigg )^{\beta} \frac{B_{\nu_1}(T)}{B_{\nu_2}(T)}
\end{equation}

We used equation \ref{eq4} to plot the flux ratio for $SPIRE$ wavebands for different $\beta$ and temperatures, as shown in figure \ref{fig:fluxRatio}. We found that the colder region seems to peak at $\beta$ $\sim$ 2 and the warm regions are more accurately defined by $\beta$ $\sim$ 1.6, giving a value of $\beta$ $\sim$ 1.8 as an intermediate value that can be used to explain the whole cloud. We have bracketed the extremes of the data with $\beta$ of 1.0 and 2.5 models.  The black error cross represents the typical flux ratio uncertainty derived from the errors in each flux as 1-$\sigma$ uncertainty.

\begin{figure}
\centering
\includegraphics[width=3.2in]{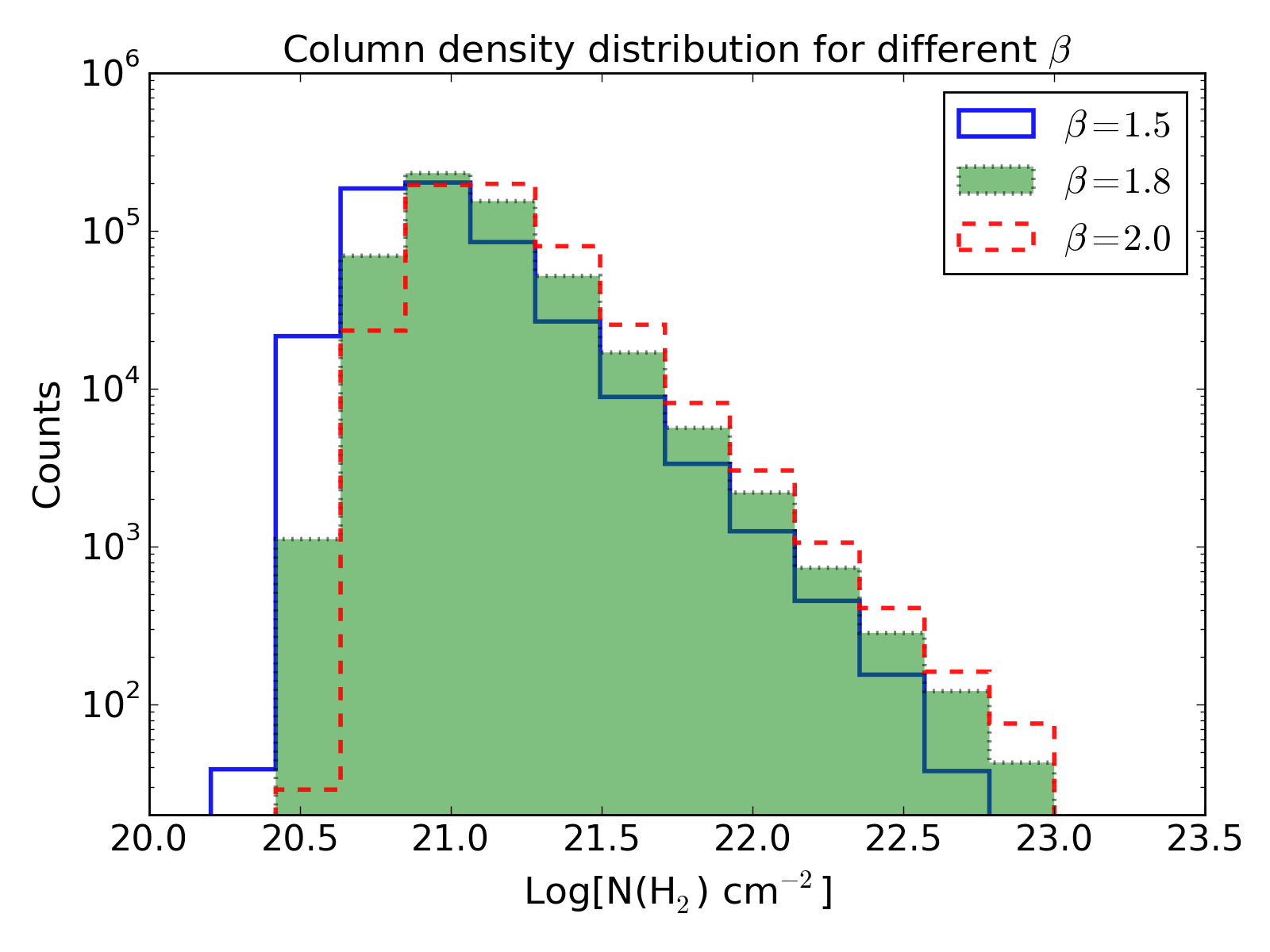}
\includegraphics[width=3.2in]{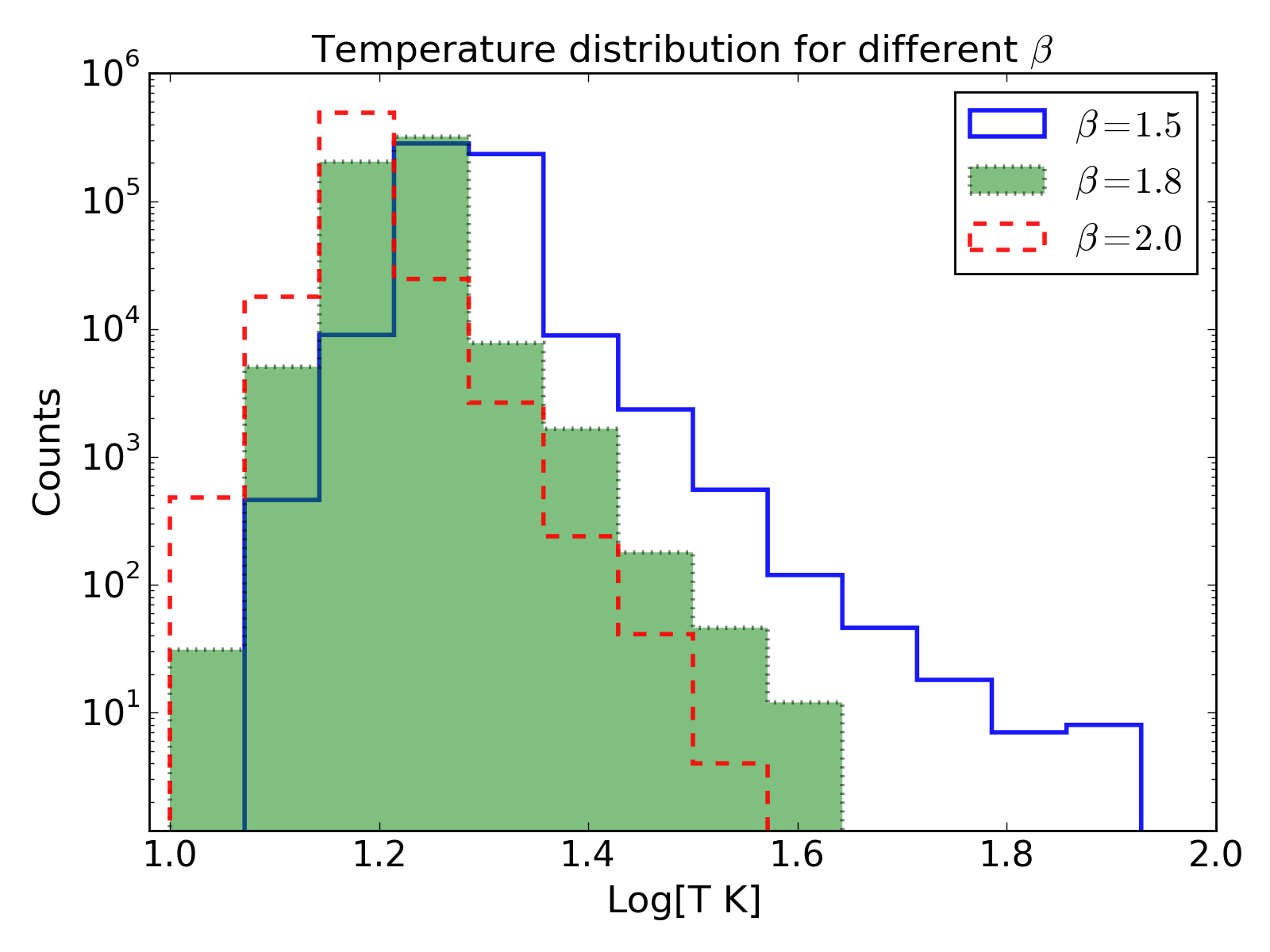}
\caption{Column density and temperature distributions in Mon~R2 after greybody fitting, for different $\beta$ values.
\label{fig:CnT}}
\end{figure}

The correct estimation of $\beta$ is crucial because the greybody fit calculations for a different value of $\beta$ can give a different estimation of temperature and column density. Thus we tested the effect of the $\beta$ uncertainty on the physical measurements that we derive from the data. Figure \ref{fig:CnT} shows the distribution of N(H$_2$) and the temperature for $\beta$ = 1.8. It also includes the possible fluctuation of $\beta$ from 1.8 to 1.5 (hotter regions) or 2.0 (colder regions). We found that the values shift by 30$\%$ in those cases, relative to the values derived using $\beta$ = 1.8. The plot also shows that if we choose a higher (or lower) $\beta$ values, we will be over-determining (or under-determining) the column density and under-determining (or over-determining) the temperature.

\subsection{Modified blackbody fits}  \label{sec:colden}

After fixing $\beta$ to 1.8, we fitted the SED for each pixel position using the PYTHON program $curvefit$ which finds the best fitting parameter values for a particular model by doing an iterative least-squares comparison between data and the theoretical model. 1-$\sigma$ flux uncertainty values for each data point are accessible from the data archival pipeline. Thus, we obtained the temperature and column density maps with 1-$\sigma$ uncertainties in each. The temperature is distributed between 11 K and 43 K and the N(H$_2$) vary between 3 $\times$ 10$^{20}$ cm$^{-2}$ and 9 $\times$ 10$^{22}$ cm$^{-2}$. The uncertainty in the canonical gas-to-dust ratio is $\sim$30$\%$ and figure \ref{fig:CnT} shows the typical uncertainty in the selection of $\beta$ as $\sim$30$\%$. Similarly, the propagated uncertainties in N(H$_2$) and the temperature values when best fitted are between 0.1$\%$ and 5$\%$. This gives the overall error in the estimation of final N(H$_2$) and temperature values to be $\sim$40$\%$. The mass-weighted mean temperature is $\sim$ 17 K. We calculated the total mass of the GMC to be $4 \times 10^{4}$ M$_{\sun}$, which is calculated over 2 $\times$ 10$^3$ pc$^{2}$ projected area. This mass estimate is similar to the value previously calculated by \cite{1992PhDT.........2X}.

\begin{figure}
\centering
\includegraphics[width=3.2in]{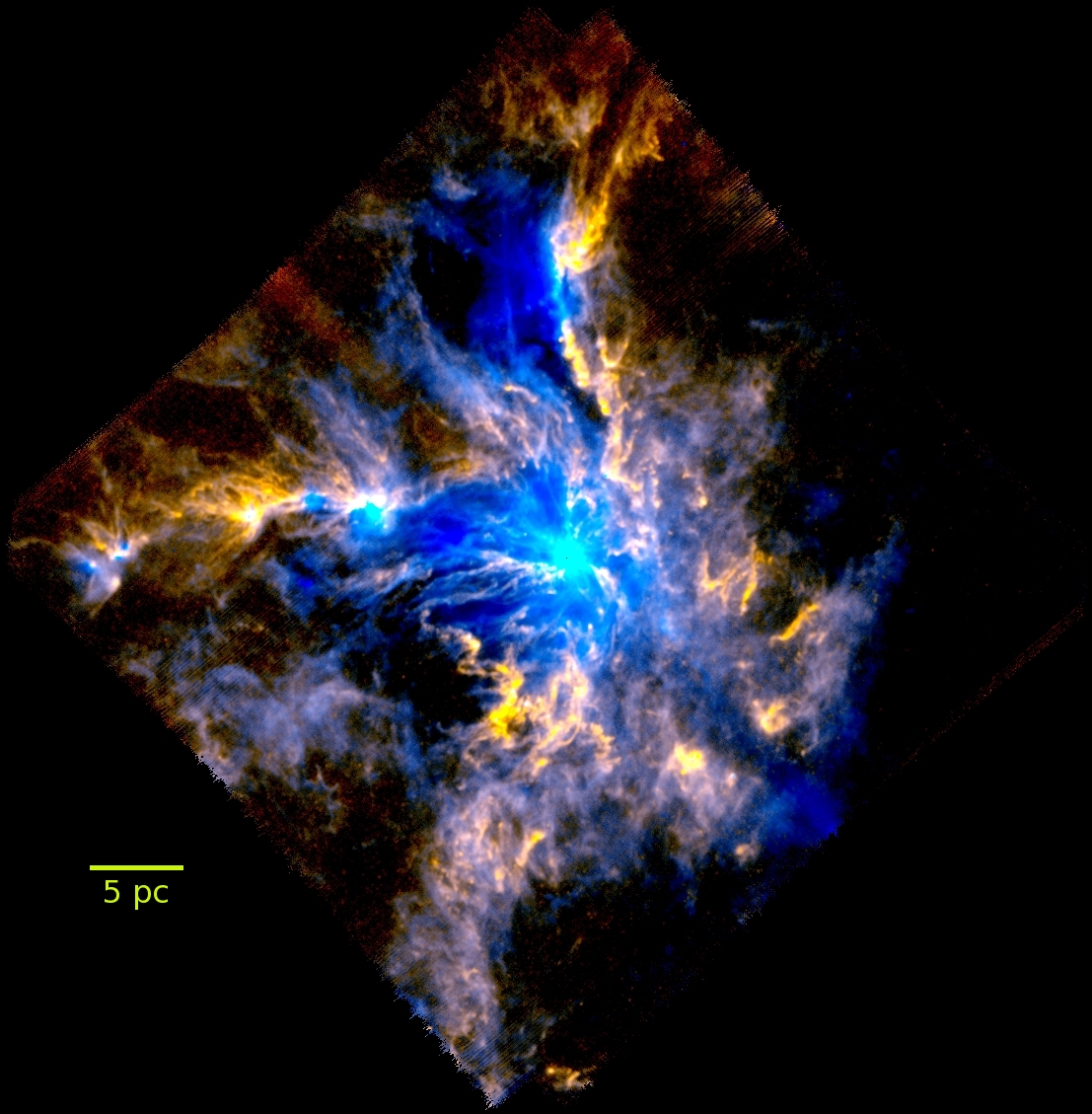}
\caption{Temperature-Column density map of Mon~R2 obtained after performing modified blackbody fits to the $SPIRE$ data.  Intensity is mapped as column density and colour is mapped as temperature where the redder areas are colder ($<$10 K) and bluer areas are hotter ($>$20 K). The typical temperature of the dust is $\sim$ 17 K.
\label{fig:monr2}}
\end{figure}

Figure \ref{fig:monr2} shows the temperature-column density map. Column density in the map is shown in terms of intensity and temperature is shown by colour, where the redder areas are colder ($<$10 K) and bluer areas are hotter ($>$20 K). Just upon visual inspection, the map clearly shows a variety of structures from diffuse regions to filaments and clumps. We can see variation in structures, from sub-parsec scales to tens of parsec scales. Some of the structures resemble elongated filaments whereas some resemble relatively round clumps, with varying column density contrast. Multiple filaments can be seen radiating towards the central Mon~R2 clump forming hub-spoke like structure \citep{2009ApJ...700.1609M}. Also filaments seem to be associated with GGD~12-15 and other clumps (c.f. figure \ref{fig:pxw}).

\subsection{Rayleigh-Jeans limitation} \label{RJlimit}

Being limited to the $SPIRE$ wavebands, our temperature dynamic range is more constrained on the high end than some comparable $Herschel$ surveys \citep{2010A&A...518L.102A}. Here, we explore the pixels for which the SED lies in the Rayleigh-Jeans (R-J) tail of the greybody spectrum. We use equation \ref{eq2} to reliably estimate the temperature and column density for the data points for which the deviation in fluxes is sufficient enough to constrain the peak of the SED. For the points where the $SPIRE$ bands fall on the R-J tail, equation \ref{eq2} can not give a reliable estimate of the parameters. Hence, here we examined the region of colour-colour space where the $SPIRE$ data would indicate that they are on the R-J regime of the SED.

\begin{figure}
\centering\includegraphics[width=3.2in]{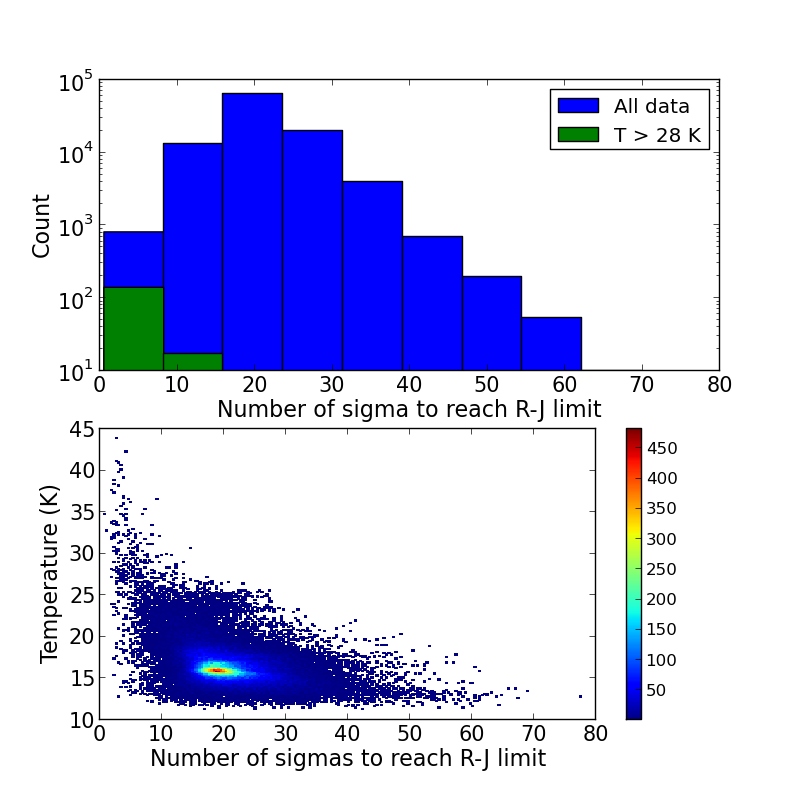}
\caption{The top panel shows the distribution of the number of sigma away from the R-J locus. Smaller numbers of sigma represent the points which are closer to the R-J locus and may be consistent with an R-J SED.  Larger numbers of sigma represent those that are clearly distinct from the R-J limit and thus are more reliably explained by the greybody emission. We overplotted another histogram for T $>$ 28 K which shows that these are the pixels with low number of sigmas. The bottom panel is a 2D histogram of the number of sigma for each pixel versus its greybody fit temperature.
\label{fig:numSig}}
\end{figure}

In figure \ref{fig:fluxRatio}, we show the loci of colours for $SPIRE$ data in the limit that the temperatures are high enough that they fall on the R-J tail. Figure \ref{fig:fluxRatio} represents a 2D histogram of actual flux ratios and the grey dashed line represents the R-J locus. The typical 1$\sigma$ uncertainty in flux ratios is represented by a black error cross on the plot, though we note that the flux ratio uncertainties vary substantially among pixels. To gauge the R-J locus proximity for each pixel, we calculated the distance to the nearest R-J point in units of sigma (cf. figure \ref{fig:fluxRatio}) and plotted them with the temperature for each pixel. Figure \ref{fig:numSig} shows the histogram of the number of sigmas required to reach the nearest point on the R-J locus and a plot of temperature with those number of sigmas. The vast majority of pixels have a large number of sigma. The high temperature pixels mostly have low number of sigma implying a higher probability of being R-J limited.

\begin{figure}
\centering
\includegraphics[width=3.2in]{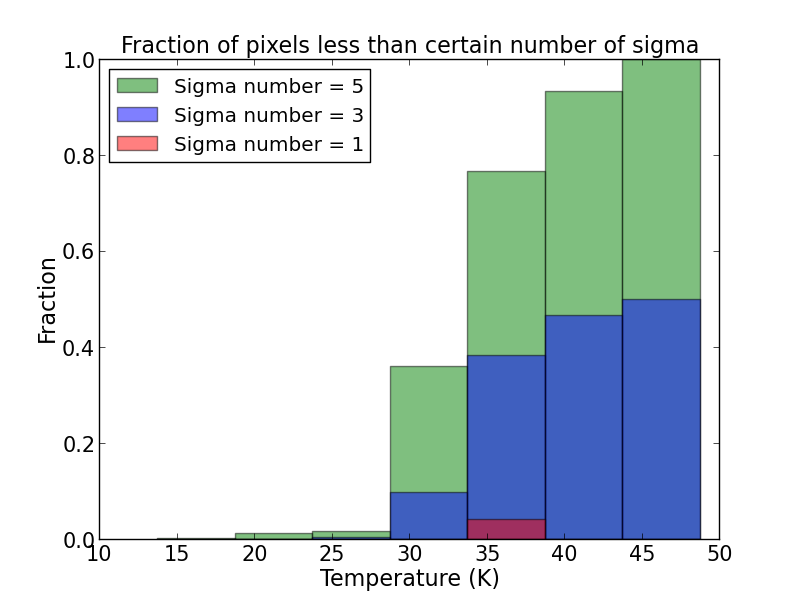}
\caption{The fraction of pixels less than a given number of sigmas vs. temperature. For temperatures greater than $\sim$28~K, the fractions begin to increase sharply. Thus, we adopt 28~K as a boundary to separate the pixels with emission lying near the R-J regime from those that follow greybody emission.
\label{fig:fracTemp}}
\end{figure}

We want to estimate the temperature threshold where the emission could be warm enough to be consistent with an R-J spectrum through the $SPIRE$ bands. For this, we binned the pixels by temperature and calculated the fraction of pixels less than a given number of sigma away from the R-J locus for each bin (see Figure \ref{fig:fracTemp}). Pixels found to be less than a few sigma have a non-negligible probability of being consistent with an R-J limited SED, and thus may lack an upper boundary on their temperature estimate. 
Thus to minimize such unconstrained fits, we picked only those pixels for which the number of sigma is greater than 5 and temperature $<$28 K for studying the column density distribution ($\S$.\ref{coldendist}). There are $\sim$500 pixels out of 10$^5$ that do not meet these requirements. We found that their exclusion doesn't significantly change the shape of the N-PDF of the whole cloud (figure \ref{fig:colden}), nor of the affected subregions ($\S$. \ref{selfsim} and figure \ref{fig:pdf_regions}).

\subsection{Emissivity calibration offset}  \label{sec:offset}

The column density values obtained by assuming a greybody emission can also be biased according to our assumed reference dust opacity. We used \citet{1994A&A...291..943O} to find the reference dust opacity corresponding to a reference frequency. The paper provides theoretical estimates of the dust opacities. Using dust emission maps we simultaneously calculate the column density and temperature. In contrast, the column densities obtained using extinction maps are temperature independent, providing a valuable check of our fits to the dust emission.

\begin{figure}
\centering\includegraphics[width=3.4in]{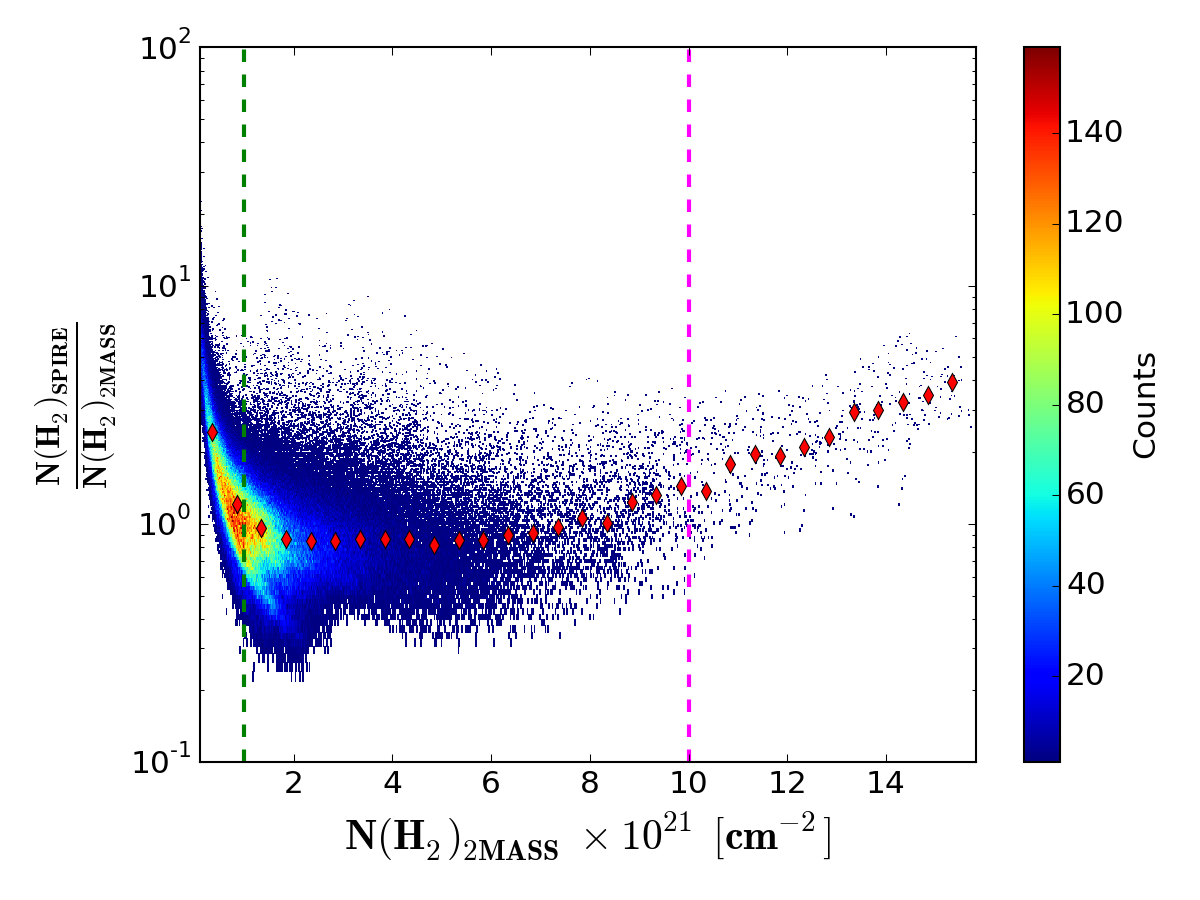}
\caption{N(H$_2$) values obtained from a $2MASS$-derived extinction map vs. the ratio of N(H$_2$) values obtained from $SPIRE$ dust emission map to the N(H$_2$) values obtained from the extinction map. N(H$_2$) values obtained from $2MASS$ data have a lower limit of 10$^{21}$ cm$^{-2}$ as the extinction values below 1 A$_{\rm{v}}$ are consistent with background noise, shown by the green dashed line. Beyond the dashed magenta line, the N(H$_2$) values from the $2MASS$ data are becoming saturated, as the high opacity of clouds obscure most background stars. Within the reliable zone enclosed by two dashed lines, the median ratio is $\sim$0.95. This consistency in the column density values obtained by two different methods gives additional confidence in our calibration assumptions.
\label{fig:offset}}
\end{figure}

\cite{2011ApJ...739...84G} used the near-IR extinction of background stars to map the dust distribution in the Mon~R2 cloud over an area similar to our $Herschel$ maps. They used $2MASS$ photometry of background stars for this purpose. We regridded the extinction maps to cover the same area and pixel positions as in the $Herschel$ maps. Figure \ref{fig:offset} shows the ratio of N(H$_2$) values obtained from $SPIRE$ emission maps to those obtained from the $2MASS$ extinction map, assuming the conversion factor of N(H$_2$) = 0.94 $\times$ 10$^{21}$ A$_{\rm{v}}$ \citep{1978ApJ...224..132B}, vs. the $2MASS$-derived N(H$_2$) values. The red diamonds represent the median ratio for each column density bin. The green dashed line represents the lower limit of the reliability zone for the $2MASS$-derived N(H$_2$) values.  Below this line, the values are consistent with noise. Similarly, the magenta dashed line represents the higher limit for $2MASS$-derived N(H$_2$).  Above this limit, the opacity of dust obscures most background stars, effectively saturating the extinction map. In the reliable zone between theese limits, the median offset in N(H$_2$) values obtained from two different methods is $\sim$5$\%$. Thus, $\sim$95$\%$ of the N(H$_2$) values are rather consistent, and our dust opacity calibration assumptions appear to yield N(H$_2$) values that are consistent with another commonly used technique.




\section{Column density distribution} \label{coldendist}

\subsection{Column density distribution function, N-PDF} \label{colden}

\begin{figure}
\centering
\includegraphics[width=3.2in]{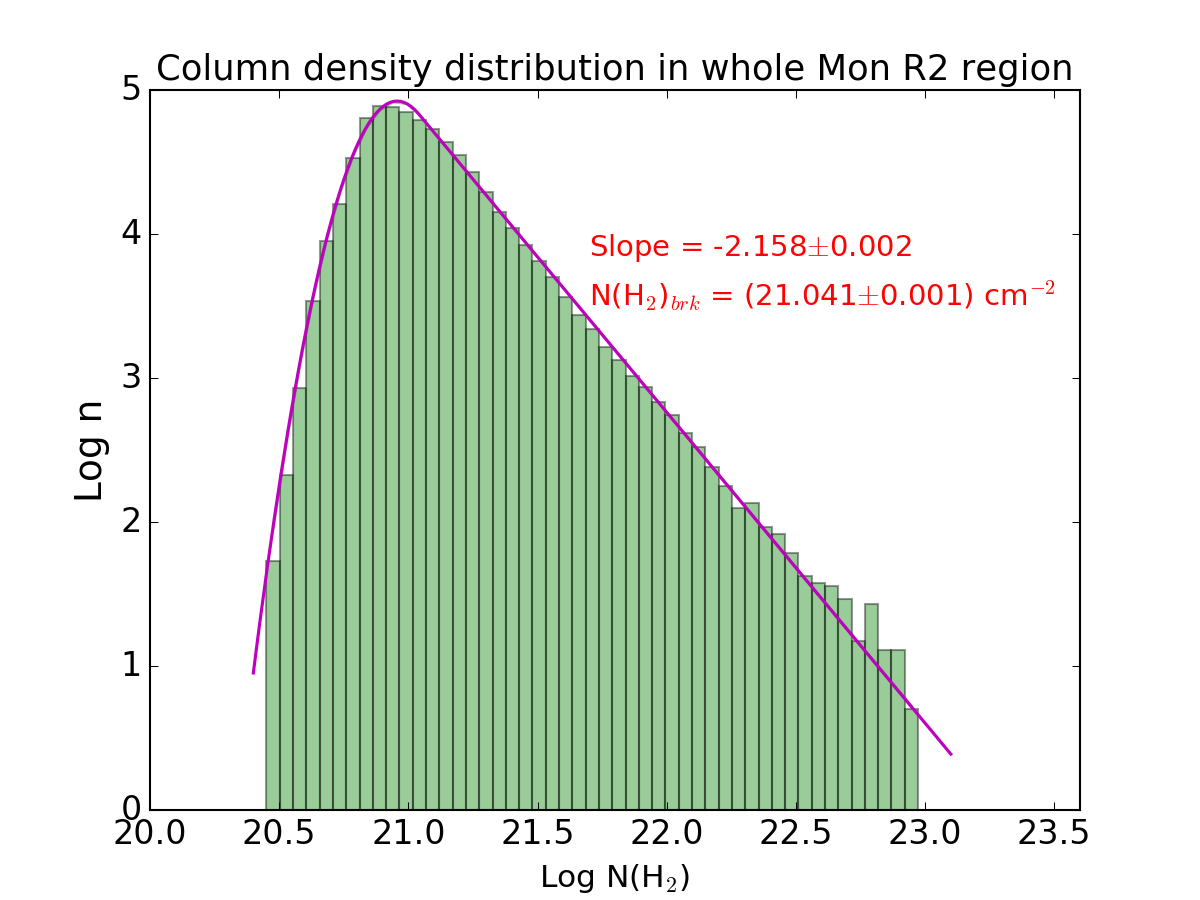}
\caption{The N-PDF of the entire Mon~R2 GMC. A lognormal distribution is seen for Log $N(H_2)$ $<$ (21.041 $\pm$ 0.001) cm$^{-2}$ pertaining to the region dominated by supersonic turbulence.  A power law nature is seen for values greater than this value with a power law index (-2.158 $\pm$ 0.002) pertaining to the regions dominated by self-gravity of gas. The y-axis represents the log of the number of 14$''$ $\times$ 14$''$ pixels.
\label{fig:colden}}
\end{figure}

Many cloud simulations predict a lognormal distribution of column densities for a cloud dominated by turbulence whereas a power law is expected to emerge when gas self-gravity wins over turbulence \citep{2000ApJ...535..869K, 2008ApJ...688L..79F}. Recently this has been simulated and studied by \cite{2014arXiv1406.4148L}. These two different natures of the probability distribution function (PDF) have been commonly observed.  However, \cite{2011ApJ...727L..20K} recently put forward the idea that the power law nature is due to regions that collapse under self-gravity and the density profiles of collapsing regions determine the power law exponent. \cite{2014arXiv1406.4148L} checked this suggestion by plotting the PDF of the regions undergoing gravitational collapse, regions largely unaffected by the gravity of the stars, and the entire simulation region before and after turning on the gas self-gravity. In their simulation, a star refers to the sink particle which is formed at a grid point at which the Jeans length falls below four grid cells \citep{1997ApJ...489L.179T}. They found that the density PDF of the non-collapsing regions matches the PDF of the entire region before inclusion of gravity.  In other words, the PDF was lognormal without a power law tail at high densities.  The implication was that regions that do not undergo collapse retain the character of pure supersonic turbulence whereas the density PDFs of collapsing regions develop a clear power law at high density. The importance of self-gravity was realised by considering several scenarios of gravitational interaction in the molecular cloud: self-gravity of gas on gas, self-gravity of stars on stars and the gravity between gas and stars. \cite{2014arXiv1406.4148L} demonstrated that the gravity due to stars does not have a significant effect on the star formation rate and gas self-gravity is the only dominant mechanism. 

While N-PDFs have been recognized as a powerful analysis tool, it is important to note that the observed shapes of N-PDFs can be impacted by effects other than gas physics.  Beam smoothing of complex projected gas geometries are the largest potential concern. In particular, small, closely spaced, high density features embedded within lower density surroundings that are smoothed by a large beam can result in substantial shifting of low density pixels upward and high density pixels downward.  Generally, the reduction of the densities of the less numerous high column density pixels will have a stronger impact on the high column density portion of the N-PDF.  Our high data quality and emphasis on relative differences in the regional N-PDF of one cloud should minimize the impact of these effects on our broader analysis, however.

\begin{figure}
\centering\includegraphics[width=3.2in]{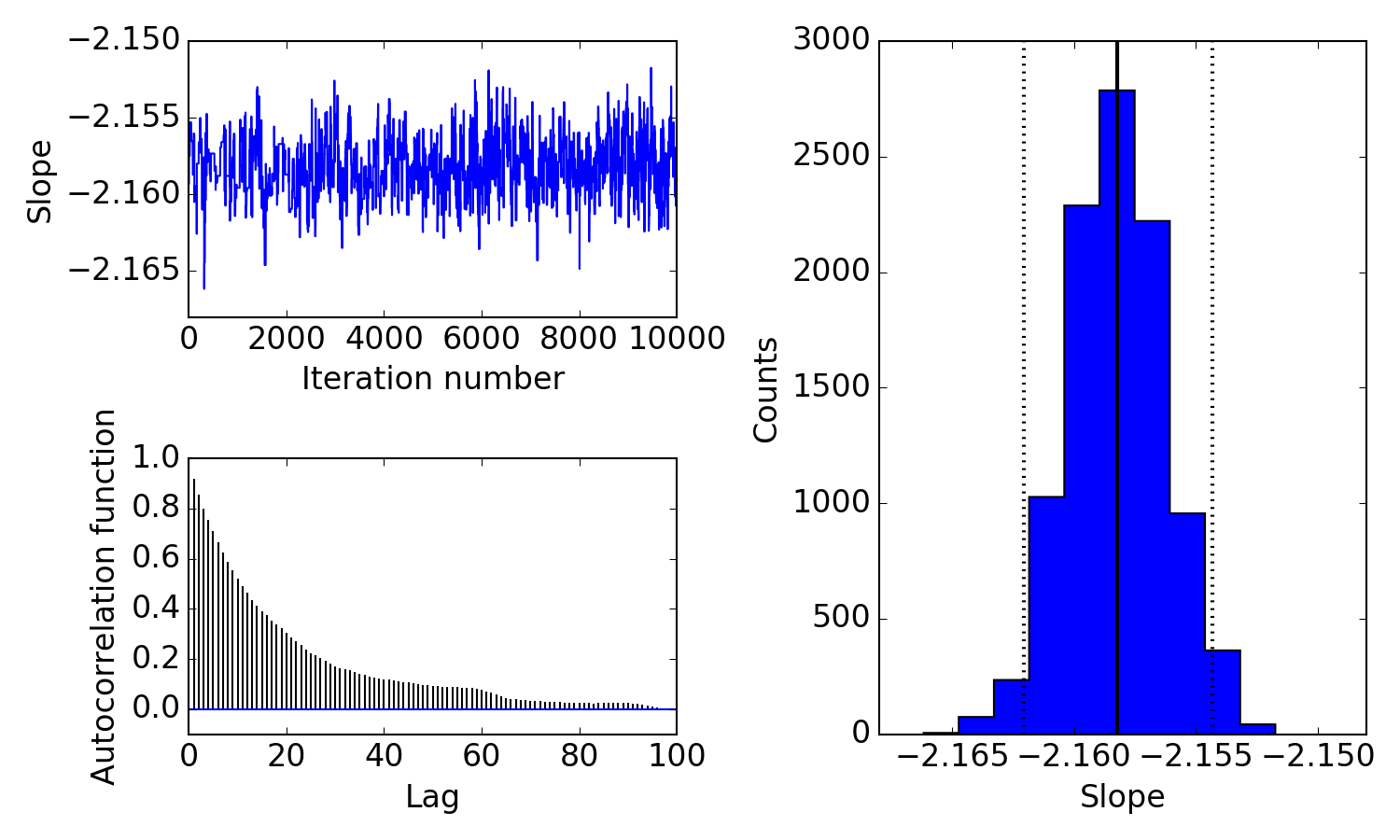}
\caption{Example plots of MCMC convergence verification. Upper left: Fluctuation of the best fitted slope in later iterations. The range of fluctuation is very small. Lower left: Autocorrelation degree of best fitted slope values. In a longer time gap, the autocorrelation goes to zero showing that the initial and final parameter values are not related. Right: Histogram of the best fitted slope values with the peak value marked by a dark black line. Black dotted lines represent the values within a 1-$\sigma$ limit.
\label{fig:gof}}
\end{figure}

Given the ongoing star formation in the MonR2 cloud, we expected to see both components, i.e., self-gravitating structures embedded within a larger, diffuse, turbulent region within a given projected region \citep{2009A&A...508L..35K}.  However, along with these two possibilities of a pure lognormal function and a combination of a lognormal and a power law function model, we found it necessary to consider a third model.  Upon carefully studying the column density distributions of several subregions (see $\S$\ref{selfsim}, below), the column density distributions of some seem to have two power laws instead of just one. The possibility of two power laws has been reported recently in other studies as well \citep{2015arXiv150708869S}. Hence, we considered the possibility of a third model with a lognormal and two power law components.


With different possibile models for the nature of an N-PDF, we generalized our fitting process to fit for the three different empirical scenarios, a lognormal function with zero to two power laws at higher column densities. If $p(x)$ is the probability distribution function, we fit the N-PDF for the whole region as well as for selected subregions ($\S$\ref{selfsim}) using the following different models:
\\

For the pure lognormal:
\begin{equation} \label{LN}
p_{G}(x) = \rm{Log}\,\Bigg[ A\, \rm{exp}\Bigg(-{\frac{(x-x_0)^2}{2\sigma^2}}\Bigg)\Bigg]
\end{equation}
where $x$ = log($N$), $N$ is the column density. $A$ is the peak, $x_0$ is the mean and $\sigma$ is the standard deviation of the distribution in log units. We have taken the log of the lognormal function because we are fitting logarithmic data (cf. figure \ref{fig:colden}).

For the combination of a lognormal with one power law:
\begin{equation} \label{LNP}
p_{G+1}(x) = 
\begin{cases}
    p_{G}(x),& \text{if } x \leq  x_{brk1}\\
    \alpha_1 x + p_{G}(x_{brk1}),              & \text{if } x > x_{brk1}
\end{cases}
\end{equation}

$x_{brk1}$ is the value of log($N$) after which the distribution takes power law form. This value is determined by the fitter itself by using it as a free parameter. The fitter is designed to look for two different subsets of data with a breaking point and fits them with the above function, considering every $N$ value in the sample space as the breaking point. Finally $x_{brk1}$ is selected by optimizing the least squares for each of the considered data.  $\alpha_1$ is the index of the power law.  The y-intercept is constrained so that the power law function is continuous with the lognormal.

For the combination of a lognormal with two power laws:
\begin{equation} \label{LNPP}
p_{G+2}(x) = 
\begin{cases}
    p_{G}(x),& \text{if } x \leq  x_{brk1}\\
    \alpha_1 x + p_{G}(x_{brk1}),            & \text{if } x_{brk1} < x \leq x_{brk2}\\
    \alpha_2 x + p_{G+1}(x_{brk2}),              & \text{if } x > x_{brk2}   
\end{cases}
\end{equation}
Similarly, $x_{brk2}$ is the value of log($N$) from where the second power law (of index $\alpha_2$ develops. These values are also used as free parameters so that we can be unbiased in our selection of the breaking points.   
\\

We used a Monte Carlo analysis to estimate the uncertainties in each bin. We randomly sampled column density values assuming that the uncertainty in column density follows a Gaussian distribution. The spread in the values for each bin gives 1-$\sigma$ uncertainty for that particular bin. We have 7 free parameters in our models and the models themselves are the combination of different functions. Hence, we need a robust fitter that can give the best fitting values from the parameter space along with reliable uncertainties. For this, we used the Markov Chain Monte Carlo (MCMC) method \citep{2003sca..book...41V}. We have followed the Metropolis-Hastings algorithm in which the samples are selected from an arbitrary ``proposal'' distribution. These samples are kept or discarded according to the acceptance rule. The whole process is repeated until we get a ``transition'' probability function so that the algorithm can transit from one set of parameter values to a more probable set.  
Based on the transition probability where the current point depends only on the previous point but yet can still span over the whole parameter space, an ergodic chain of positions in parameter space is formed, known as the Markov Chain. The Markov Chain samples from the posterior distribution ergodically assuming the detailed balance condition. MCMC estimates the expectation of a statistic in a complex model by doing simulations that randomly select from a Markov Chain. We have used the PYTHON package $Pymc$ for this purpose \citep{2010jsc...35.1P}.

\begin{figure}
\centering
\includegraphics[width=3.0in]{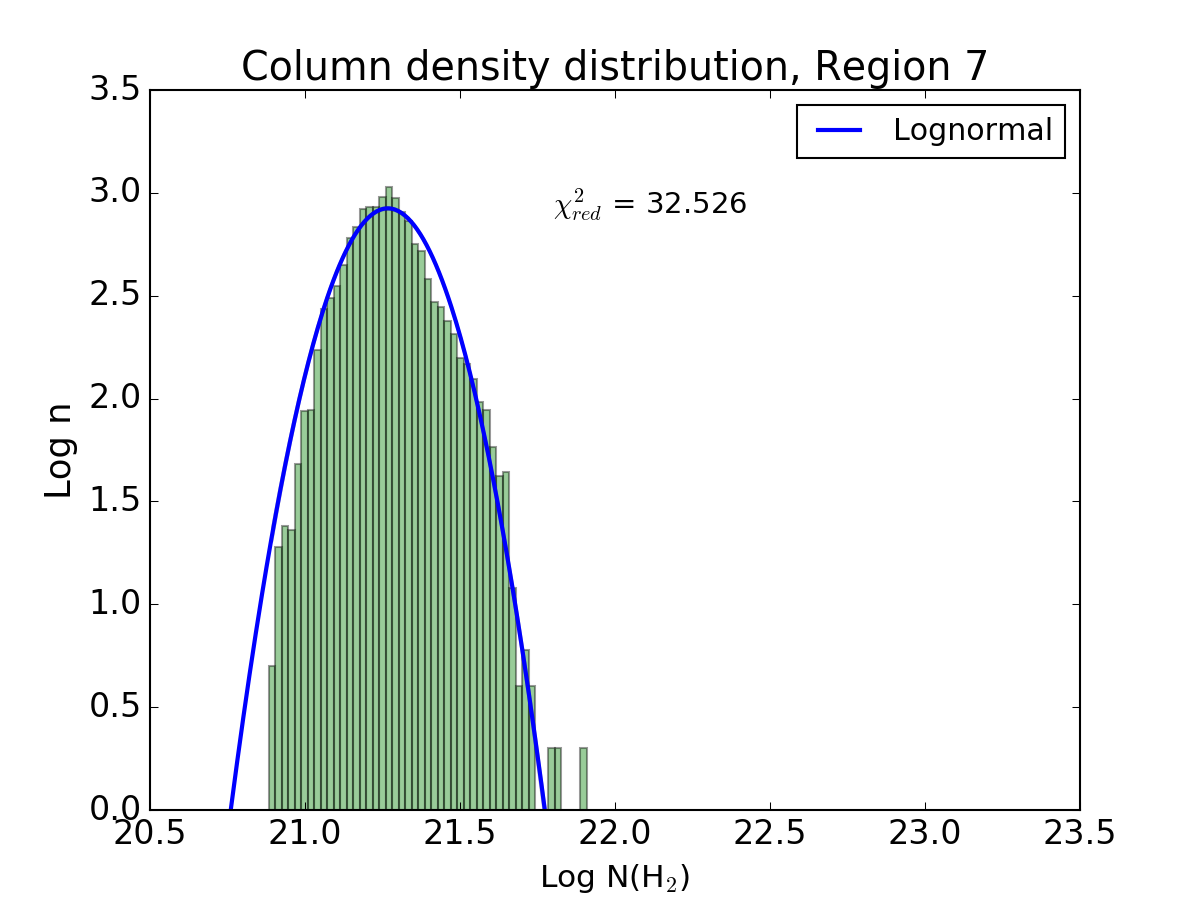}
\includegraphics[width=3.0in]{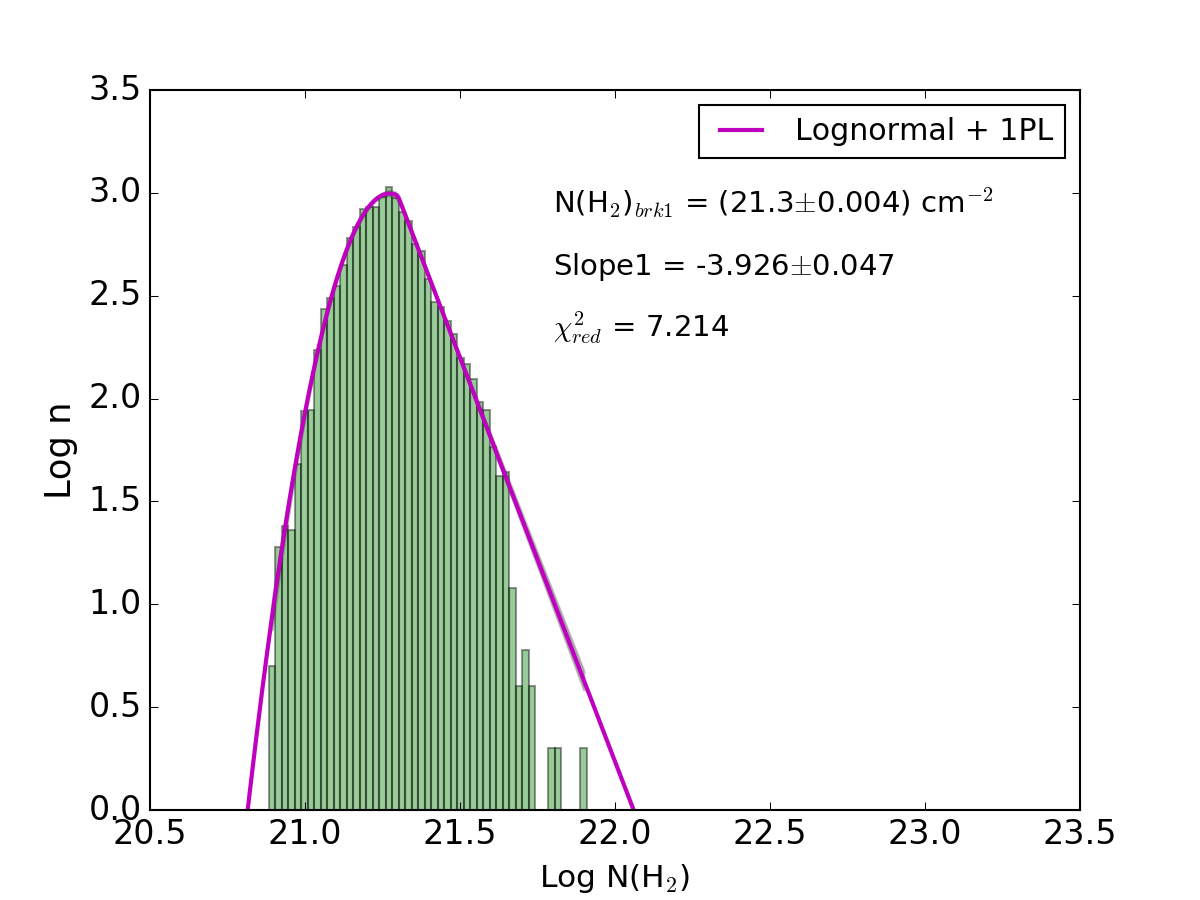}
\includegraphics[width=3.0in]{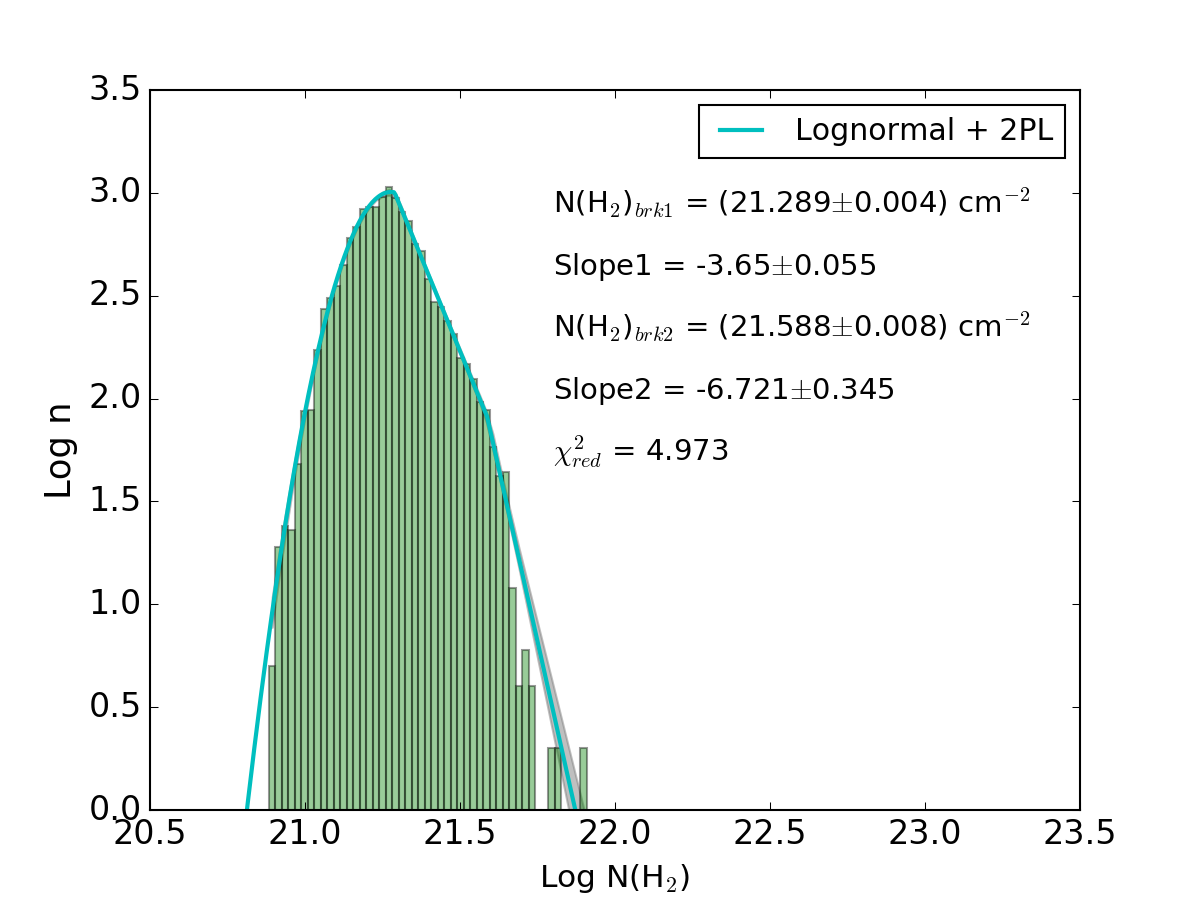}
\caption{The best fitting models overplotted on the column density distribution in region 7 for all three fitting scenarios.  The top panel shows the best fit using $p_{G}(x)$ (pure lognormal), the mid panel shows $p_{G+1}(x)$ (lognormal and one power law) and the bottom panel shows $p_{G+2}(x)$ (lognormal and two power laws). The 1-$\sigma$ error in the model fits is shown in shaded grey. `$N(H_2)_{brk1}$' is the value where the first power law tail appears (in log units) with slope of `Slope1'.  `$N(H_2)_{brk2}$' is the value where the second power law tail begins (also in log units) with slope of `Slope2'. The y-axis represents the log of the number of 14$''$ $\times$ 14$''$ pixels.
\label{fig:NPDF_reg7}}
\end{figure}

We fitted all N-PDFs using equations \ref{LN}, \ref{LNP} \& \ref{LNPP} and accepted the model that has the minimum reduced chi-square value. For the whole Mon~R2 GMC region, $p_{G+1}(x)$ with a lognormal and a single power law fits the distribution best (see figure \ref{fig:colden}).  We see the lognormal nature below a critical value of column density, log N(H$_2$) = (21.041 $\pm$ 0.001) cm$^{-2}$. The lognormal behaviour is centered at log N(H$_2$) = (20.957 $\pm$ 0.001) cm$^{-2}$, with a characteristic peak, log(n) = (4.923 $\pm$ 0.001) and width of (0.13 $\pm$ 0.001). For log N(H$_2$) $>$ (21.041 $\pm$ 0.001) cm$^{-2}$, a very prominent power law with index (-2.158 $\pm$ 0.002) emerges. We note a caveat that the N(H$_2$) values derived from $Herschel$ data for A$_K< 0.1$, or N(H$_2$) $\lesssim$ 10$^{21}$ cm$^{-2}$ may not be securely determined due to large foreground/background emission superpositions \citep{2015A&A...576L...1L}. However, for our study we limit this possibility because of the location of the cloud. Furthermore, they show that the N-PDF shape below A$_K \sim 0.1$ changes according to our selection of the boundary and choice of baseline subtraction. Hence, for the remaining of this paper we are characterizing the low end of the N-PDF for the sake of completeness and we do not analyze the lognormal portion of the fit results.

In the MCMC chain, we discarded the first 50$\%$ of the iterations in the so-called ``burn-in'' period and examined the other half of the iterations to see whether the parameter values converge. The convergence of parameter values as estimated by the posterior probability distribution function is robust as we can see in figure~\ref{fig:gof}. We plotted the trace of the best fitted parameter values for an acceptable period of iteration (second half period for our case), the autocorrelation degree in the parameter values in different time lags, and invested the distribution of parameter values by plotting the histogram to check the parameter convergence to assure a strong goodness of fit. In figure \ref{fig:gof}, we have shown such plots for one of the parameter values, the slope of the power law portion of figure~\ref{fig:colden}. The plots for the other parameters are similar.

\begin{figure*}
\centering
\includegraphics[width=5.3in]{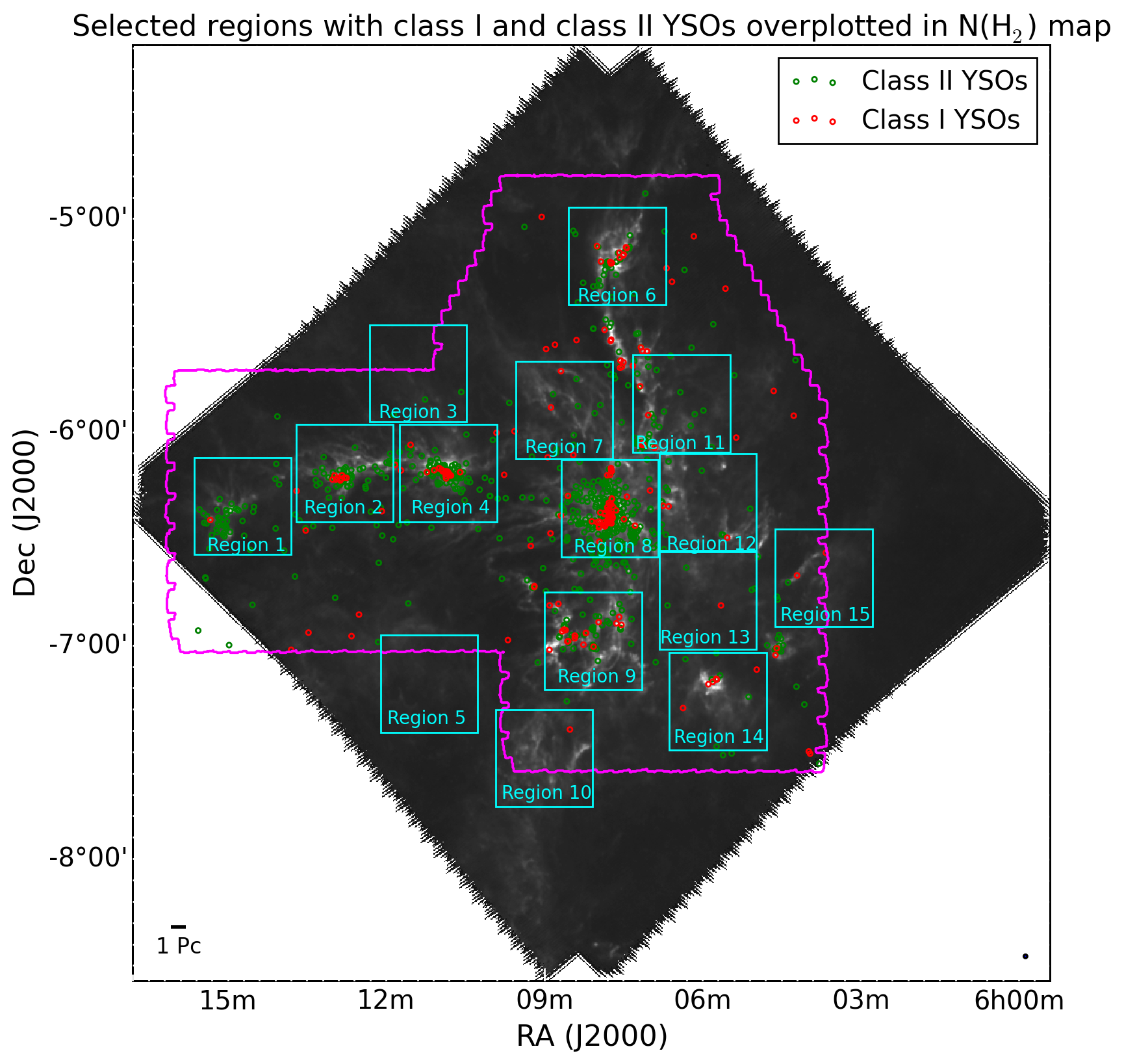}
\caption{Overlay of the 15 regions (cyan squares) for which we studied the N-PDF, plotted on the temperature-column density image of figure~\ref{fig:monr2}. The regions are defined according to the prevalence of high and low column density gas and the presence of YSOs. We have overplotted Class~I (red circles) and Class~II (green circles) YSOs along with the $IRAC$ coverage contour (magenta).
\label{fig:regions}}
\end{figure*}

The emergence of two different PDFs (lognormal and power law) for possibly two different scenarios has been invoked by recent observational studies (\citealt{2008A&A...480..785L, 2015arXiv150708869S}). Generally the clouds with active star formation show a power law tail for higher column densities along with lognormal nature for lower column densities. In contrast, almost all quiescent clouds have PDFs that are either well described by a lognormal function over the entire column density range or else they only show relatively low excess (power law tail) at high column densities \citep{2009A&A...508L..35K}. However, it has also been suggested that the low column density feature for star forming molecular clouds might well be the residuals caused by other physically distinct clouds lying along the line of sight or a maniefestation of the uncertainties in low extinction estimations \citep{2015A&A...575A..79S}.  In our case, the Mon~R2 GMC lies $\sim$11$^{\circ}$ below the galactic plane, thus such chance superpositions are unlikely, as seen in the study of molecular line data by \cite{1992PhDT.........2X}. Furthermore, our effort to cross-calibrate with {\it Planck} data has resulted in reasonable signal to noise at low column densities (Figure~\ref{fig:fluxvserr}).

\subsection{Regional N-PDF analysis} \label{selfsim}

\begin{figure*}
\centering
\includegraphics[width=2.2in]{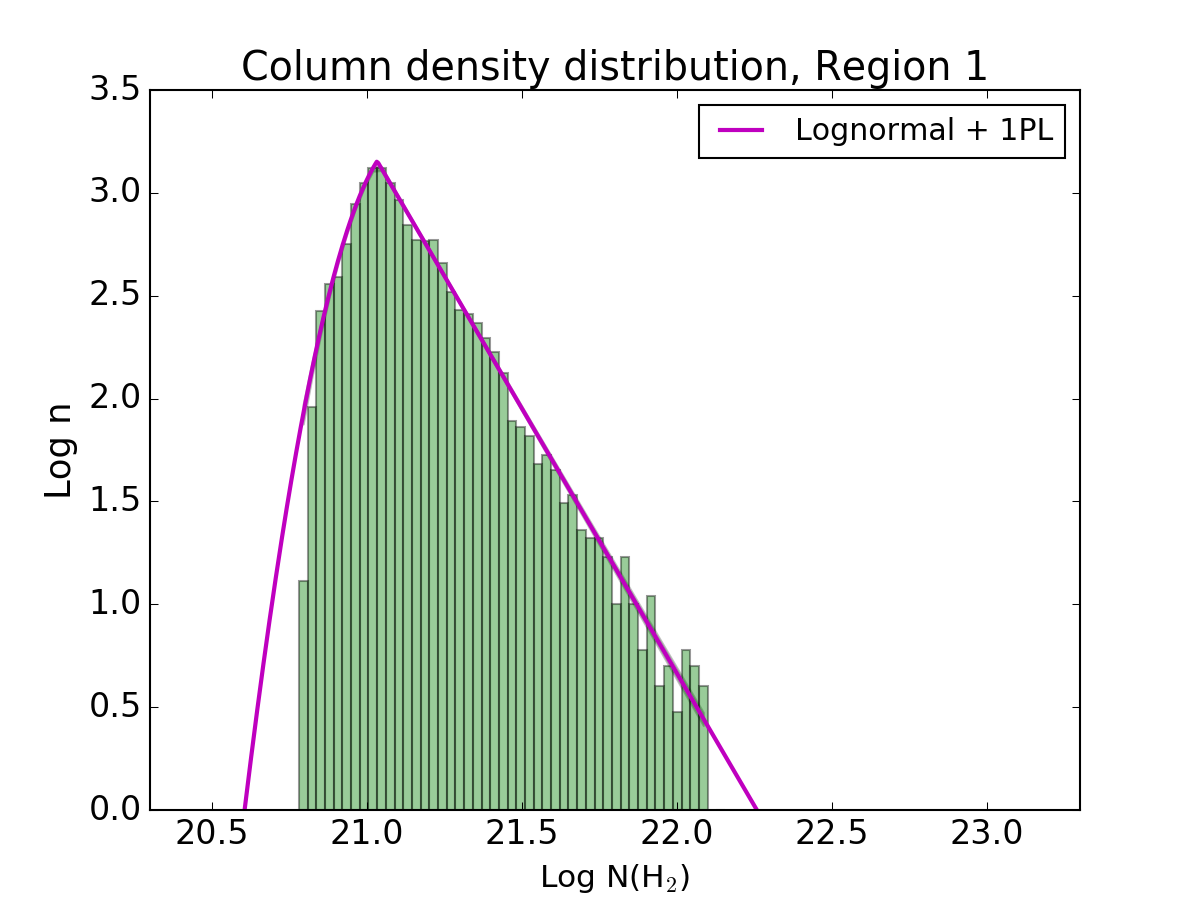}
\includegraphics[width=2.2in]{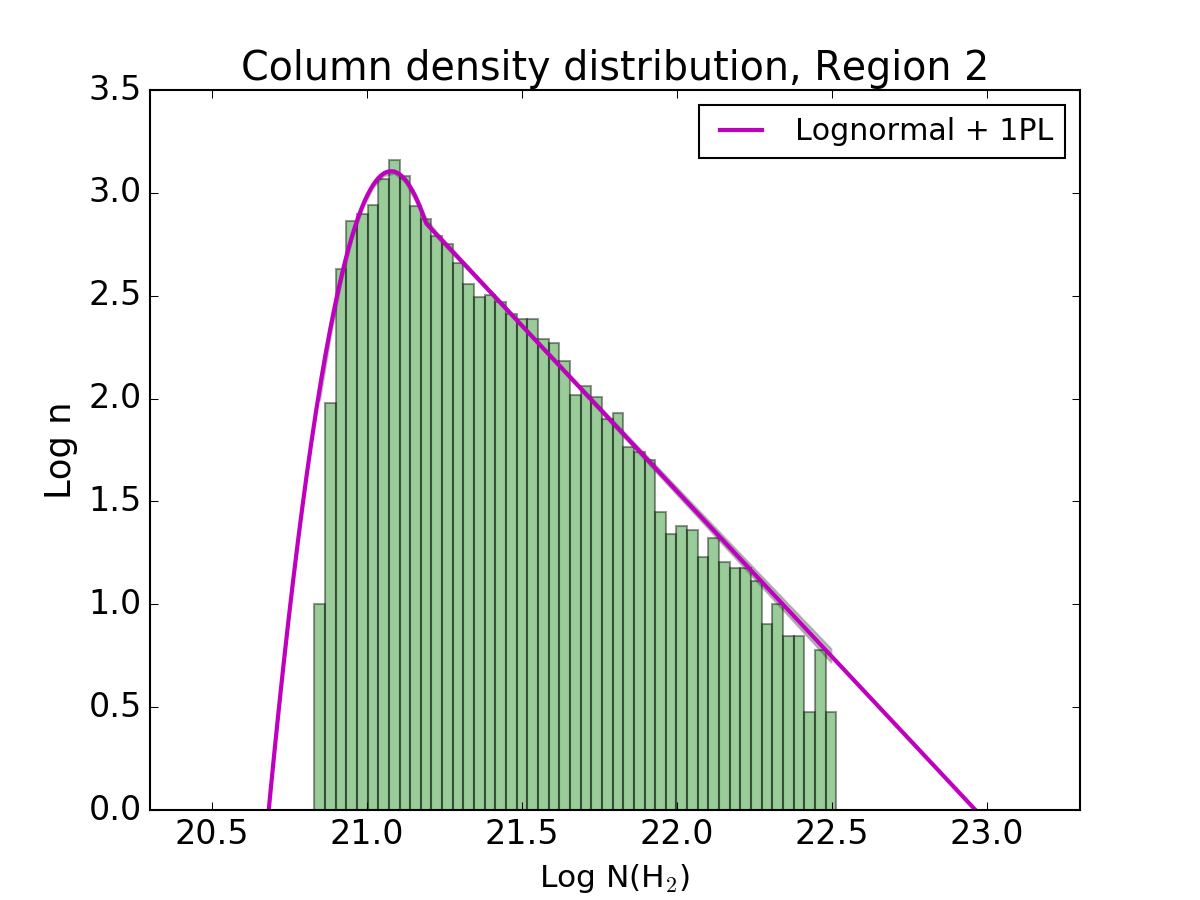}
\includegraphics[width=2.2in]{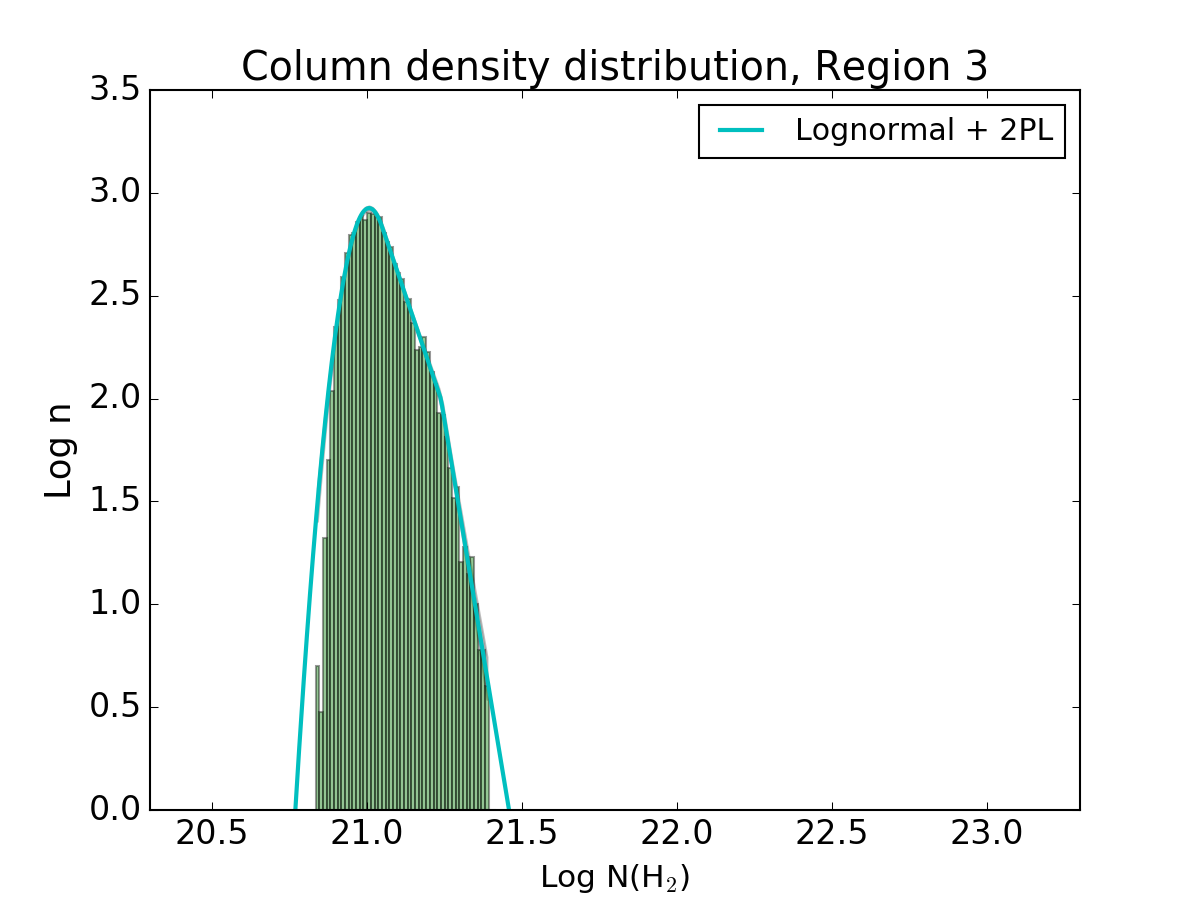}
\includegraphics[width=2.2in]{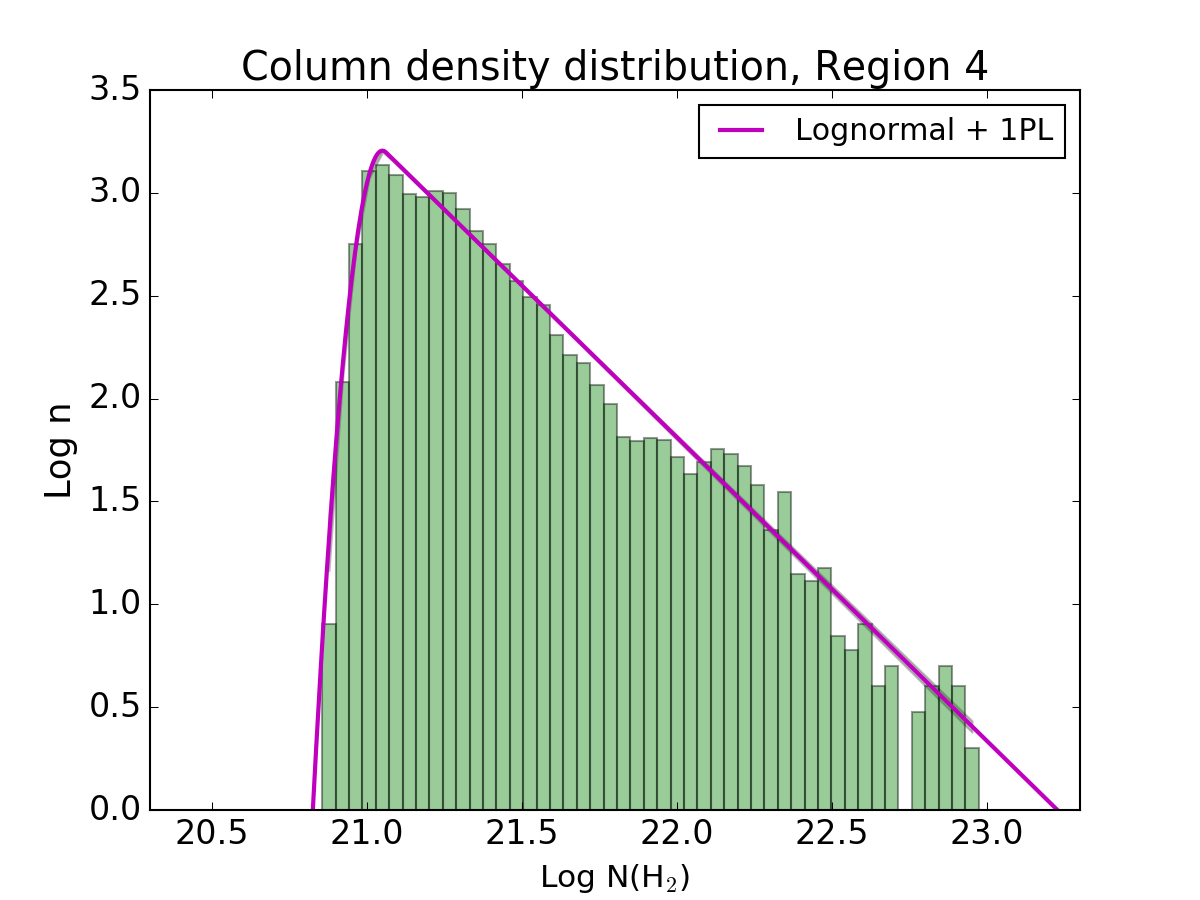}
\includegraphics[width=2.2in]{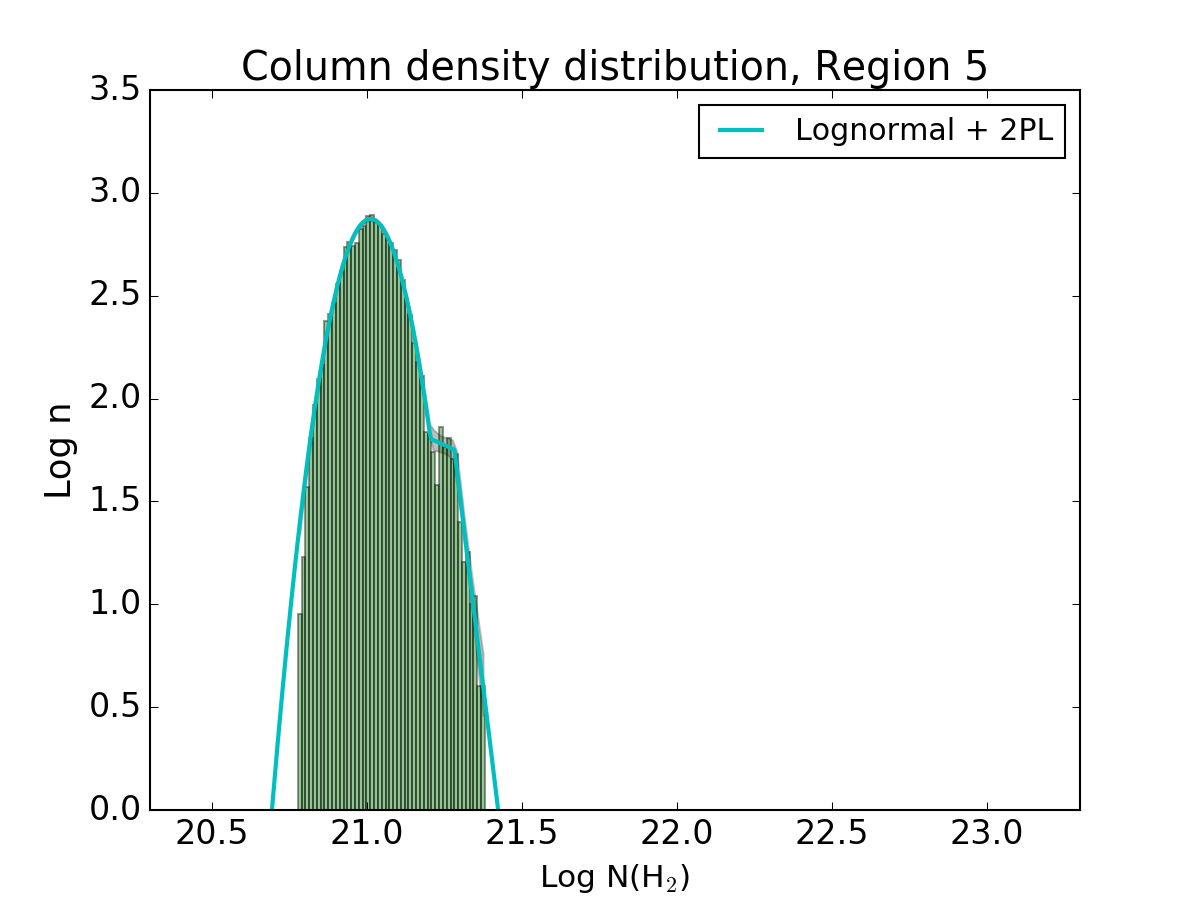}
\includegraphics[width=2.2in]{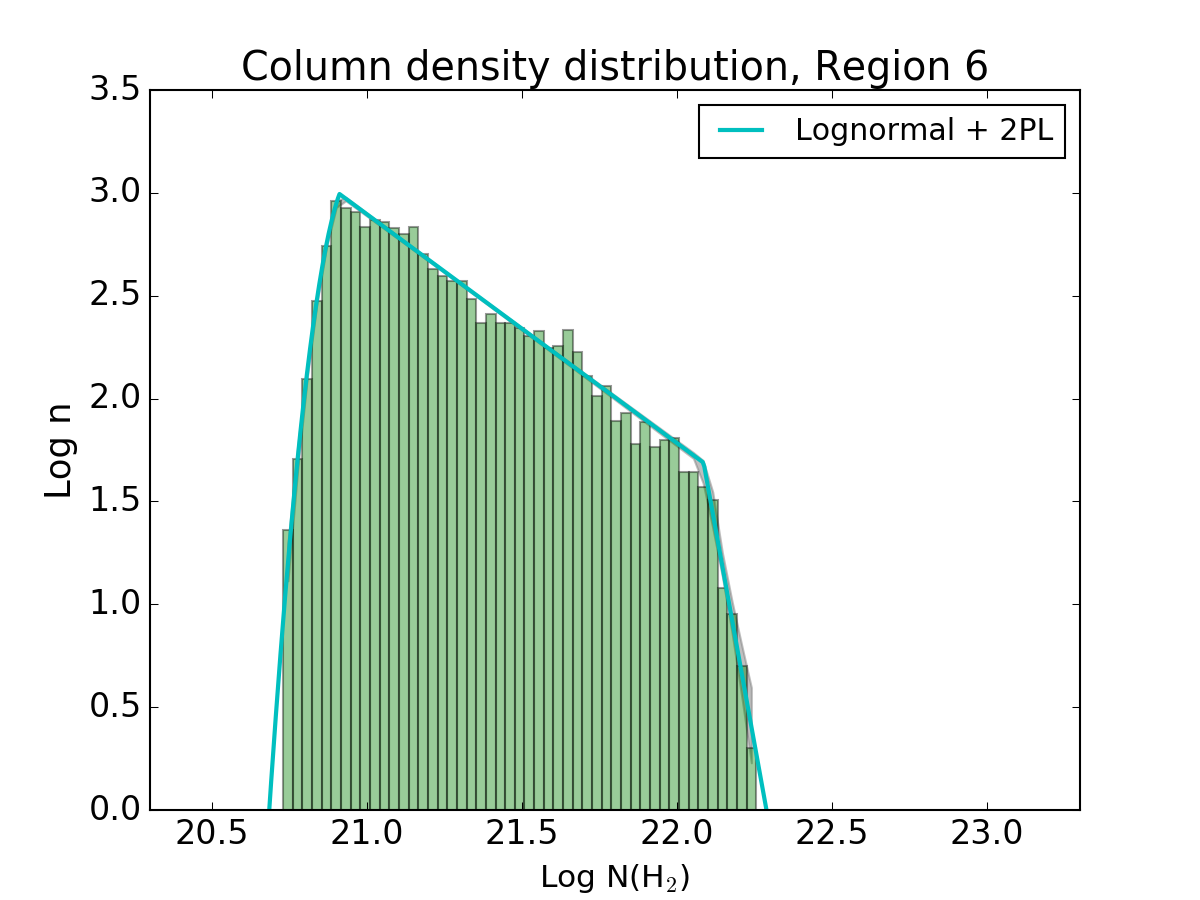}
\includegraphics[width=2.2in]{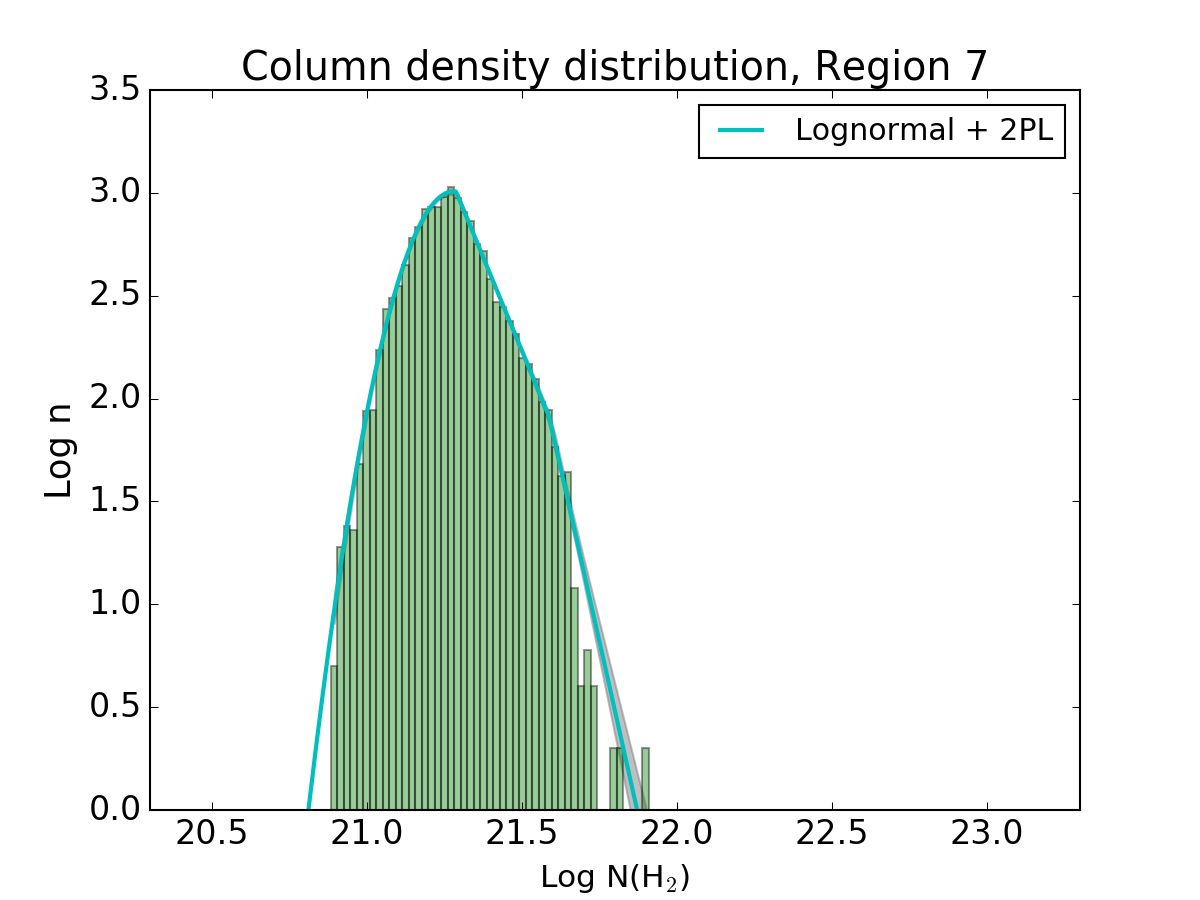}
\includegraphics[width=2.2in]{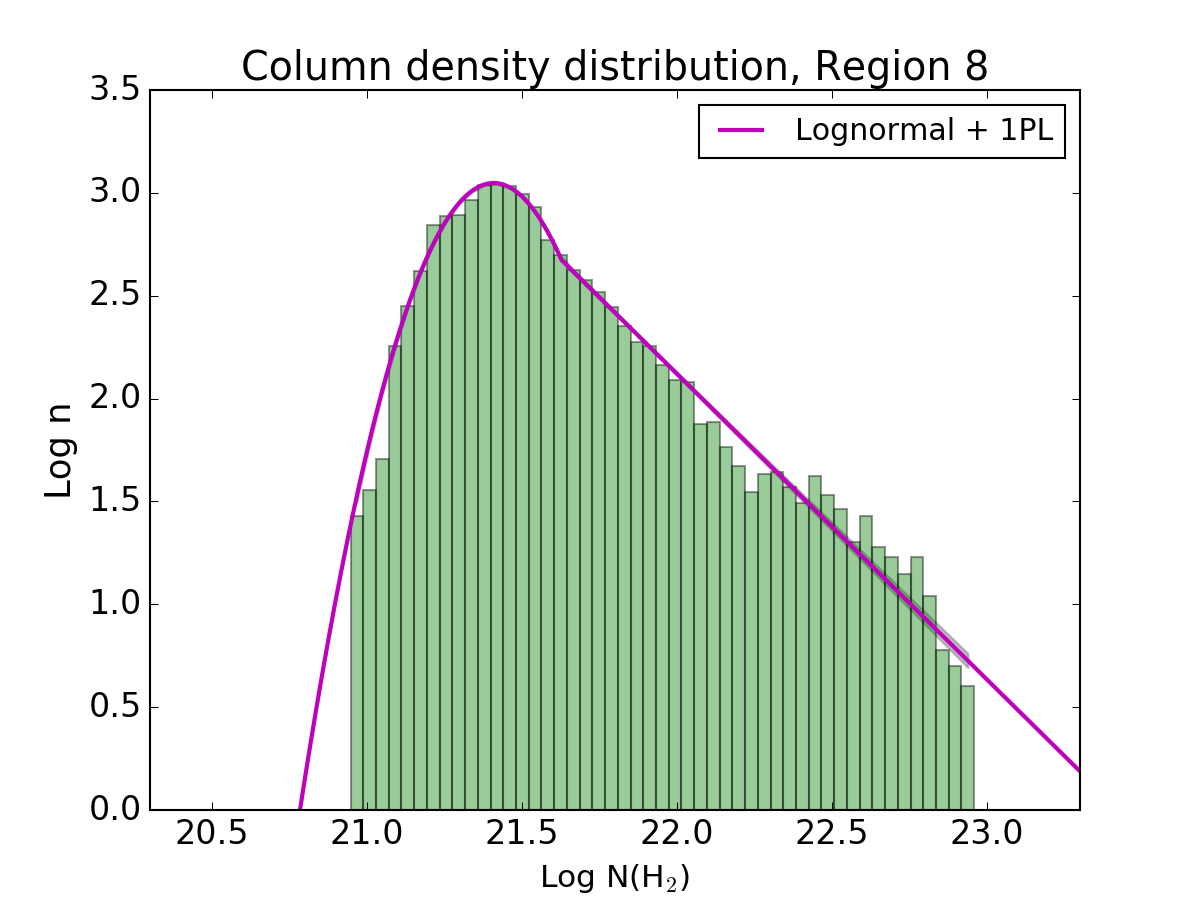}
\includegraphics[width=2.2in]{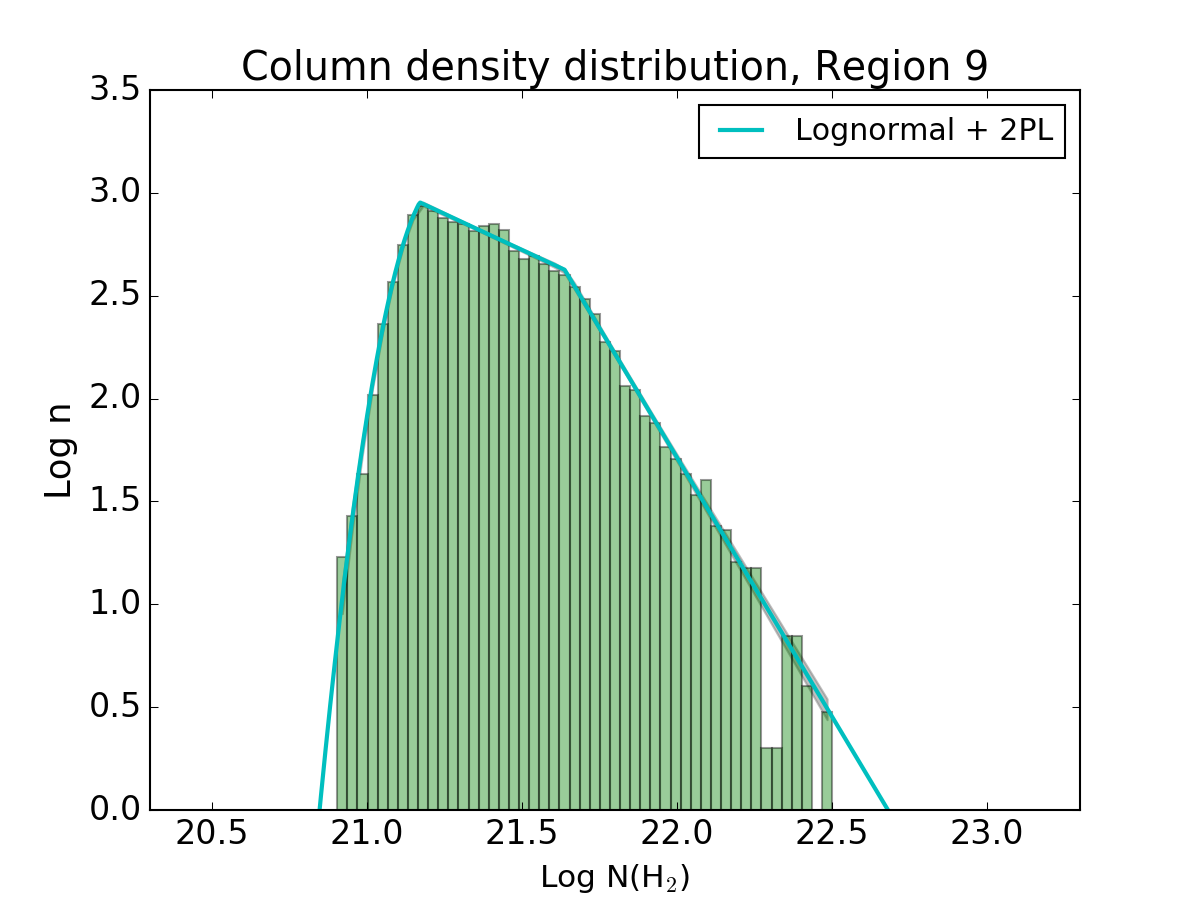}
\includegraphics[width=2.2in]{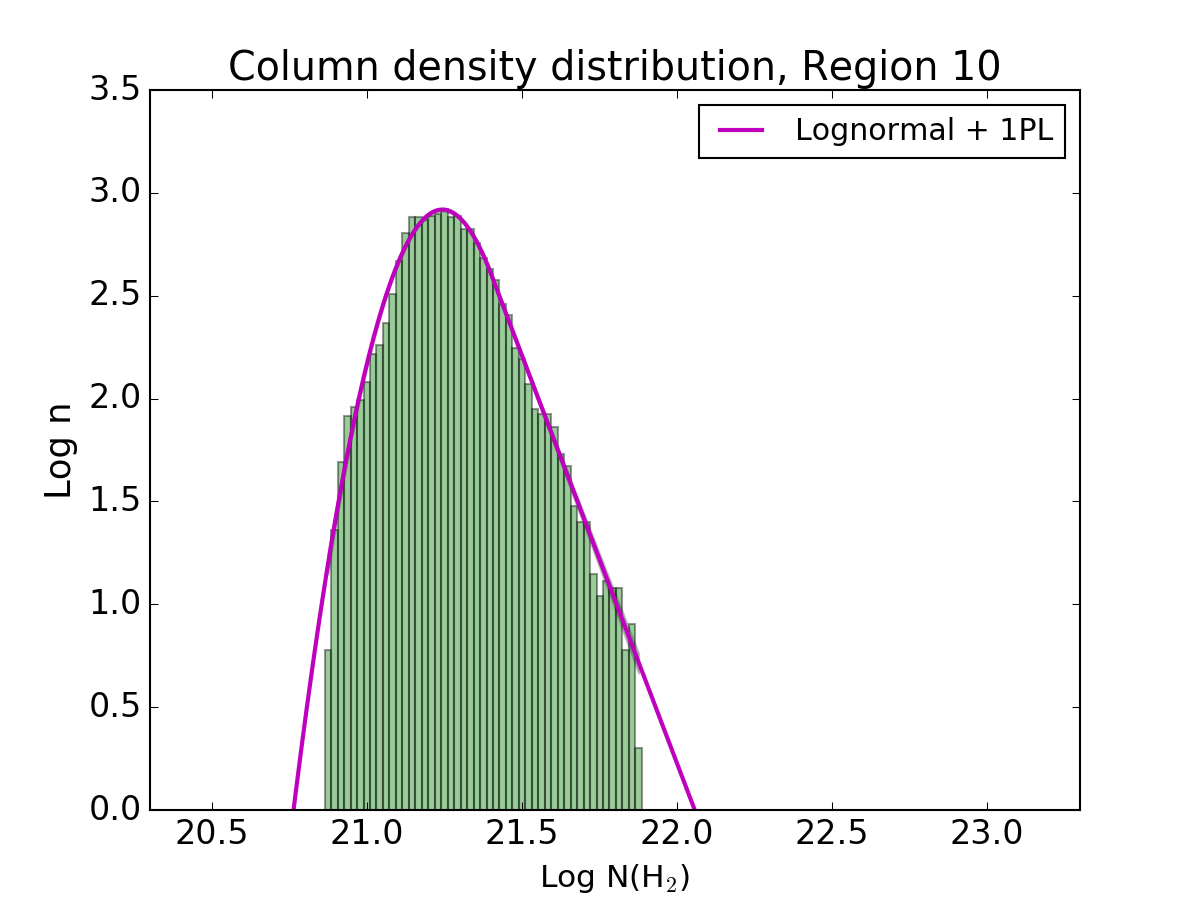}
\includegraphics[width=2.2in]{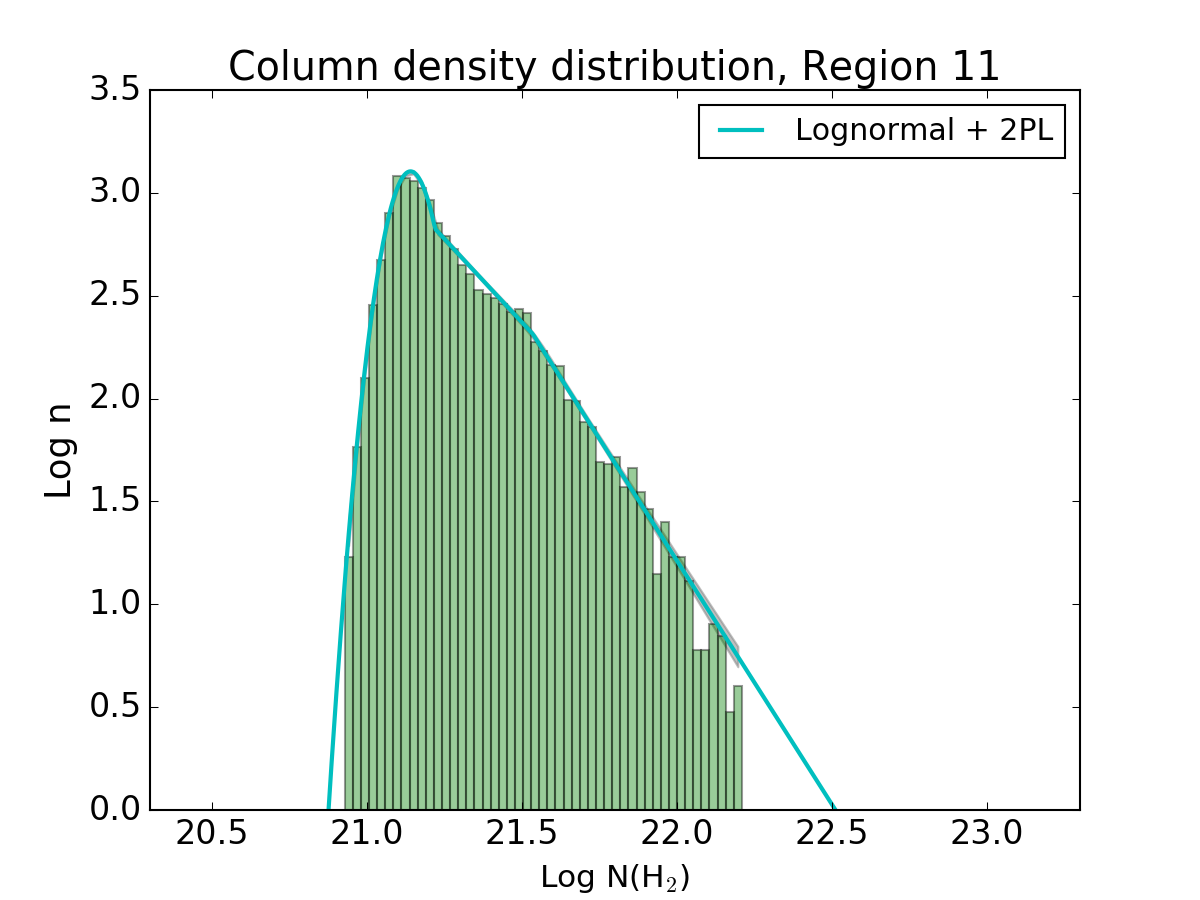}
\includegraphics[width=2.2in]{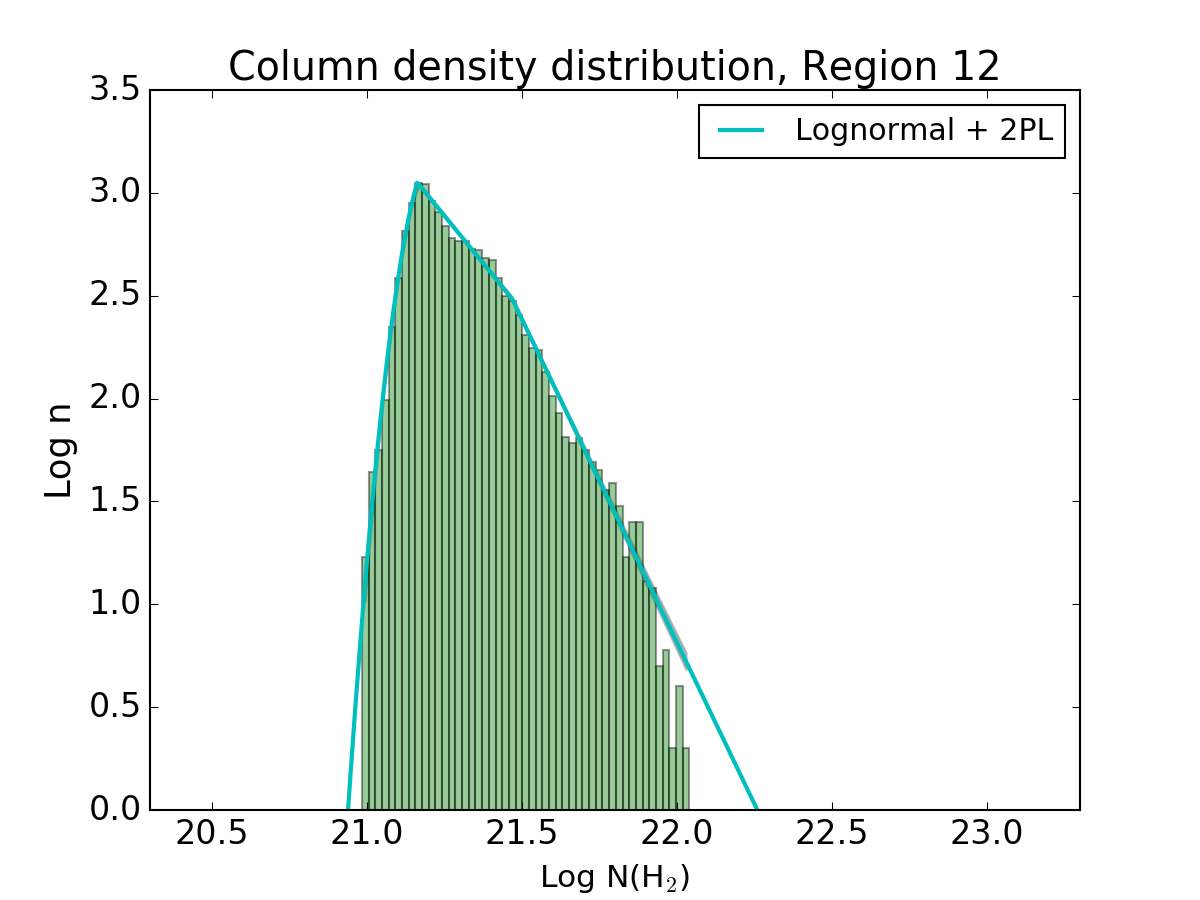}
\includegraphics[width=2.2in]{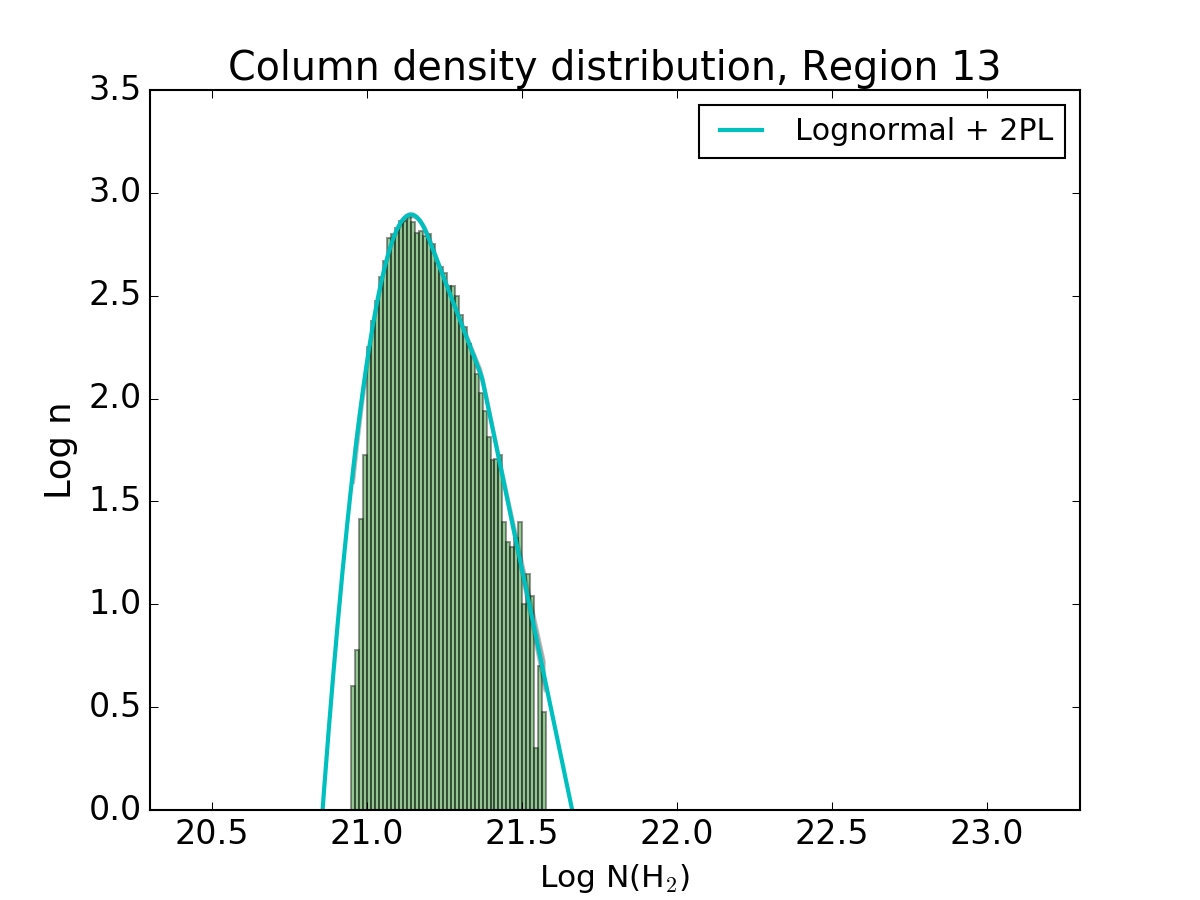}
\includegraphics[width=2.2in]{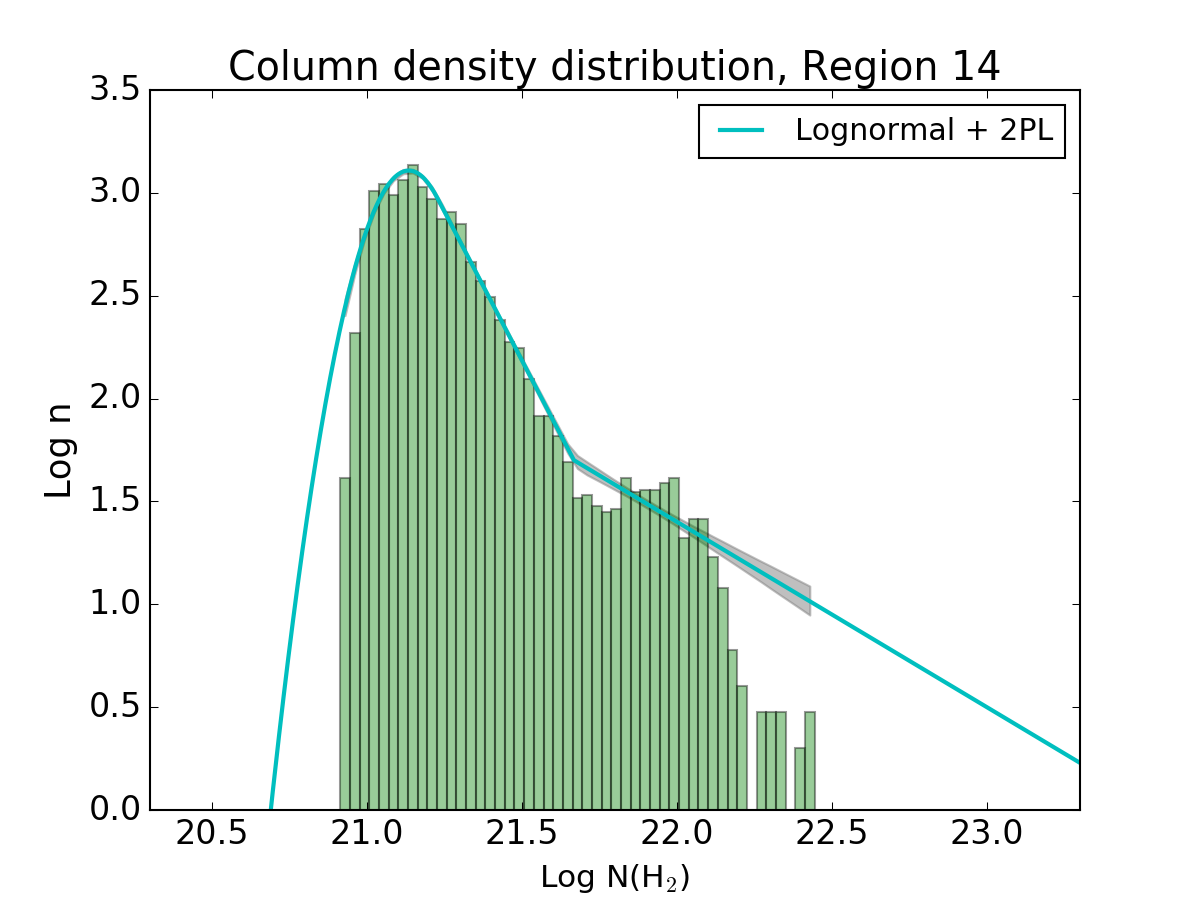}
\includegraphics[width=2.2in]{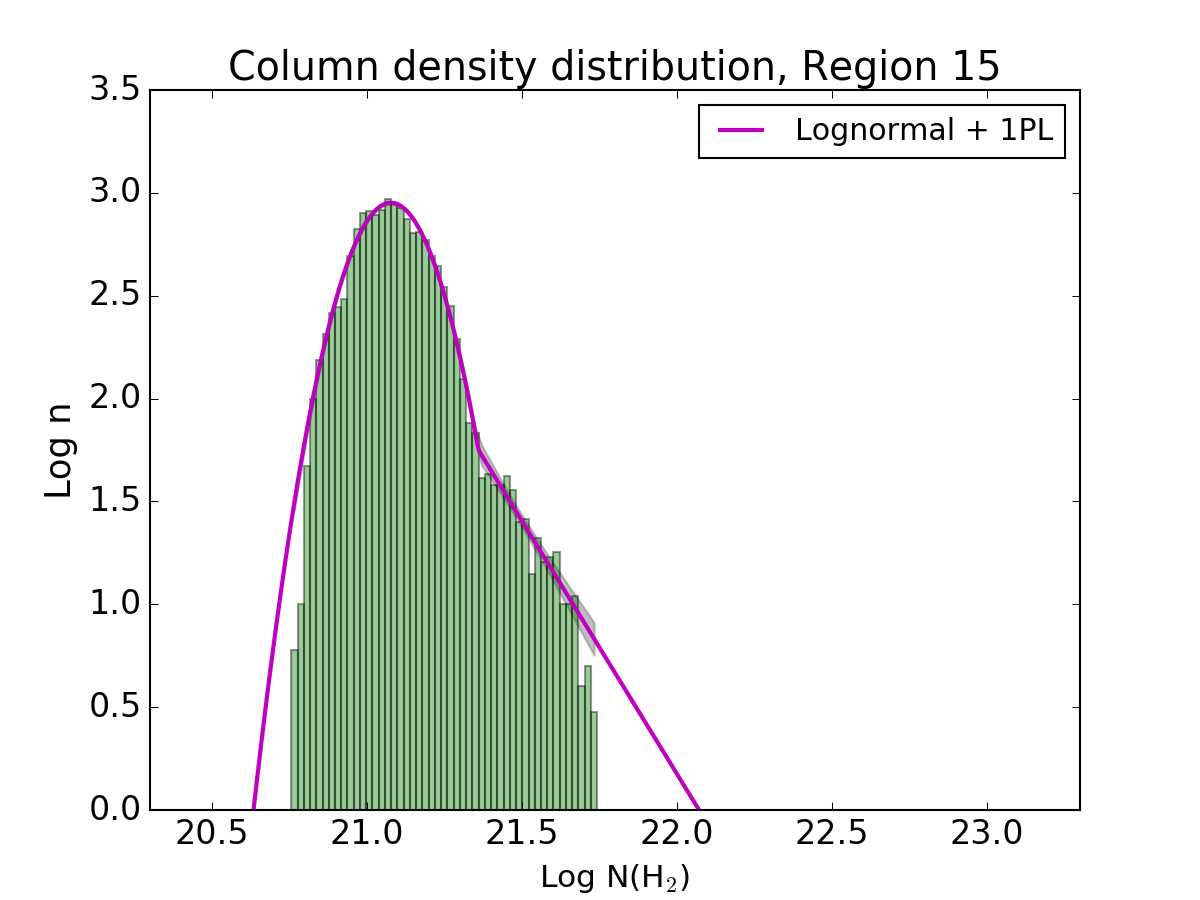}
\caption{The accepted models (c.f. table \ref{tab:chisq}) overplotted on the column density distributions for each of the region shown in figure \ref{fig:regions}. The best fitted parameter values with their uncertainties as obtained from MCMC are listed in table \ref{tab:param}. We see that the regions with stellar cluster cores in them are best fitted by the combination of lognormal and one power law model (Lognormal + 1PL). The regions that lack a central core and mostly contain over dense filamentous gas structures are described by the combination of lognormal and two power law functions (Lognormal + 2PL). The region with very few YSOs and mostly diffuse gas is described by the pure lognormal function. Some regions are poorly fit regardless of model (c.f. region 14), and this effect is quantified in the reduced $\chi^2$ in table \ref{tab:chisq}. The y-axis represents the log of the number of 14$''$ $\times$ 14$''$ pixels.
\label{fig:pdf_regions}}
\end{figure*}

In addition to studying the global characteristics of the N-PDF as shown in figure \ref{fig:colden}, we checked if similar behaviour is seen if we do the same study on smaller scale subregions within Mon~R2. Similar work has been done by \cite{2015A&A...577L...6S} in Orion~A where they find that the power law slopes of the regional N-PDFs is correlated with the fraction of Class~0 protostars in that region. For our case, we selected 15 regions in the cloud that sample the range of YSO density environments \citep{2011ApJ...739...84G}.  We want to explicitly study the N-PDF behaviour for diffuse regions with no or very little YSOs and the regions with dense YSO clusters. The selected regions are all equal area (6.5 $pc$ $\times$ 6.5 $pc$) to facilitate fair comparison at a fixed size scale.

The possible models are: the lognormal function (equation \ref{LN}, $G$), the combination of a lognormal and a single power law function (equation \ref{LNP}, $G+1$) and the combination of a lognormal and two power law functions (equation \ref{LNPP}, $G+2$). We fitted each region's N-PDF with all three models and computed the resulting reduced $\chi^2$ values. As an example, the best fitting plot using all three models for region 7 are shown in figure \ref{fig:NPDF_reg7}. All other regions are treated similarly. The acceptance of a particular model is set based on the reduced $\chi^2$ values. Theoretically, for a good fit, the reduced $\chi^2$ value should be close to $\sim$1.  For each region, the accepted model is the one whose reduced $\chi^2$ value doesn't decrease by more than 20$\%$ while going from simplest to the most complex model. Table \ref{tab:chisq} contains all of the reduced $\chi^2$ values. The values for the accepted models are highlighted in bold. The best fitted parameter values with their uncertainties as estimated using MCMC for the accepted models are shown in table \ref{tab:param}.

\begin{table}
  \begin{tabular}{cccc}
  	\hline
    Region & $\chi^2_{Lognormal}$ & $\chi^2_{Lognormal + 1PL}$ & $\chi^2_{Lognormal + 2PL}$ \\ 
    \hline
1	&	104.46	&	\bf{7.34}	&	7.68 \\
2	&	83.36	&	\bf{12.82}	&	11.52 \\
3	&	19.20	&	6.46	&	\bf{3.82} \\
4	&	102.03	&	\bf{31.79}	&	31.37 \\
5	&	16.47	&	4.60	&	\bf{2.75} \\
6	&	66.37	&	17.98	&	\bf{11.80} \\
7	&	32.53	&	7.22	&	\bf{4.98 }\\
8	&	245.28	&	\bf{13.53}	&	11.68 \\
9	&	59.90	&	45.70	&	\bf{5.12} \\
10	&	34.38	&	\bf{6.66}	&	6.97 \\
11	&	96.56	&	8.50	&	\bf{6.35} \\
12	&	82.52	&	15.12	&	\bf{6.12} \\
13	&	18.17	&	8.91	&	\bf{6.80} \\
14	&	129.26	&	44.91	&	\bf{26.35} \\
15	&	53.27	&	\bf{5.54}	&	7.33 \\
	\hline
  \end{tabular}
  \caption{Reduced $\chi^2$ value for the three models that we considered. From the simple to complex models: lognormal; lognormal and one power law; lognormal and two power laws. If the reduced $\chi^2$ value in a more complex model doesn't decrease by more than 20$\%$, we accept the simpler model as the representative model. The values corresponding to the accepted model according to this acceptance rule are shown in bold font.}
  \label{tab:chisq}
\end{table}


Figure \ref{fig:regions} shows our adopted sub-division of the Mon~R2 column density map into 15 different regions and we have overplotted Class I and class II YSOs on the map along with the $Spitzer$-$IRAC$ mid-IR coverage contour from \cite{2011ApJ...739...84G}. The accepted model is overlayed on the N-PDF for each region in figure \ref{fig:pdf_regions}. We did not find a region with minimum reduced $\chi^2$ for a $G$ model (equation \ref{LN}) as the most favourable model. We expect a pure lognormal behavior for non-star-forming turbulent gas, whereas all of our regions have at least a few YSOs, except region 5 where we have incomplete $Spitzer$ sampling. Even regions chosen for their low gas density and dearth of YSOs exhibit some small N-PDF excess above the lognormal. The regions that are best fitted by a $G+1$ model (equation \ref{LNP}) include all of the well-known embedded YSO clusters (e.g. the central Mon~R2 cluster is in region 8, GGD~17 in region 2, and GGD~12-15 in region 4; \citealt{2009ApJS..184...18G}) centered on visually obvious gas ``hubs'' with filamentary structures radiating outward \citep{2009ApJ...700.1609M}. Finally, the $G+2$ model (equation \ref{LNPP}) regions visually appear to be aggregates of several distinct filamentary gas structures with no dominant dense gas hubs, as seen in region 6, 7, 9, 11 and 12. Detailed characterizations of the radial profiles of gas filaments in nearby clouds show a spatially resolved maximum in the column density along the spine of the filaments (e.g. \citealt{2011A&A...529L...6A}). Thus, in the N-PDFs of regions which consist of aggregate of filaments, we see a downturn of the N-PDF near the maximum density of the most massive filament. These regions correspond to relatively diffuse and young YSO distributions, and will be discussed further below.



To understand how the star formation properties depend on the gas properties, we compare the number of sources to both the power law exponent and the mass, all of which are tabulated in Table \ref{tab:starGasProp}. For studying the variation of the power law index with the YSO count, we have considered the power law index from both $G+1$ and $G+2$ models. For the N-PDFs with two power laws, we use the first index because the second power law index generally seems to represent a cut-off at some near-maximal N(H$_2$) value, likely set by the peak column density of the densest filament in the aggregate. The total gas mass is calculated by integrating masses over every positive pixel value for each region. \cite{2015A&A...576L...1L} have called into question the robustness of $Herschel$ derived N(H$_2$) values below A$_K \sim 0.1$ mag ($\sim$10$^{21}$~cm$^{-2}$). The upper limit to the effect this may have on our masses is given by a hypothetical box of 6.5 $\times$ 6.5 pc with a mean extinction of 0.1 A$_K$; this box has a total mass of $\sim$673 M$_{\odot}$. This is a substantial fraction of the reported total mass in the regions with low mean N(H$_2$), such as regions 3 and 5. Finally, YSOs are counted by the data presented in \cite{2011ApJ...739...84G} for each of the 15 regions; we only include YSOs superimposed on gas column densities in excess of $N(H_2) > 10^{22}$~cm$^{-2}$ (e.g. \citealt{2014ApJ...780..173B}). We note that a few regions (3, 5, 10, 15) are not fully surveyed by Spitzer, however the preponderance of low column density gas in the missed areas suggests that at most a few YSOs would be omitted from those regions \citep{2011ApJ...739...84G}.

\begin{figure}
\centering
\includegraphics[width=3.2in]{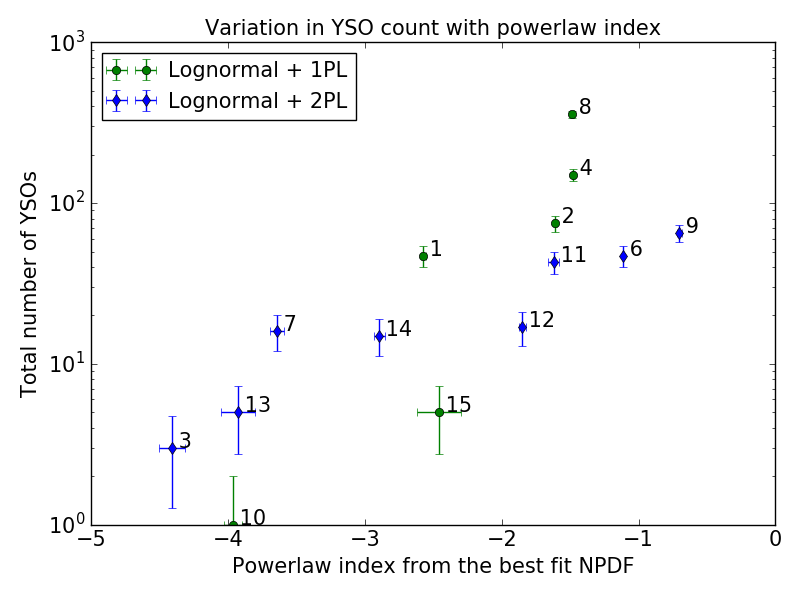}
\caption{Total number of YSOs vs. first power law index of the best fitted N-PDF model by region. Uncertainties in YSO count follow Poisson statistics whereas the power law index uncertainies are obtained from Markov Chain Monte Carlo fits. The numbers associated with the data points refer to the region number (cf. figure \ref{fig:regions}). A steep correlation is seen for the regions that are defined by $G + 1$ models. The dependence is present, but shallower, for the regions defined by $G + 2$ models.
\label{fig:plVsyso}}
\end{figure}

In previous studies, it has been shown that the exponent of the power law component relative to the lognormal component of an N-PDF correlates with more active star formation \citep{2009A&A...508L..35K}. We find this same correlation within the Mon R2 cloud. In figure \ref{fig:plVsyso}, we see a correlation between YSO count and the first power law index ($\alpha_1$) when these data are separated by the N-PDF model type (Pearson coefficient 0.91 and 0.93 for $G+1$ and $G+2$ types, respectively). The correlation is steeper in the case of the single power law models, while it is shallower for those with two power laws. Since the single law power laws PDFs have a higher fraction of gas mass at higher column densities, the higher number of YSOs in these regions may result from a higher star formation rate density and more efficient star formation at these high gas densities \citep{2011ApJ...739...84G}.

\begin{figure}
\centering\includegraphics[width=3.2in]{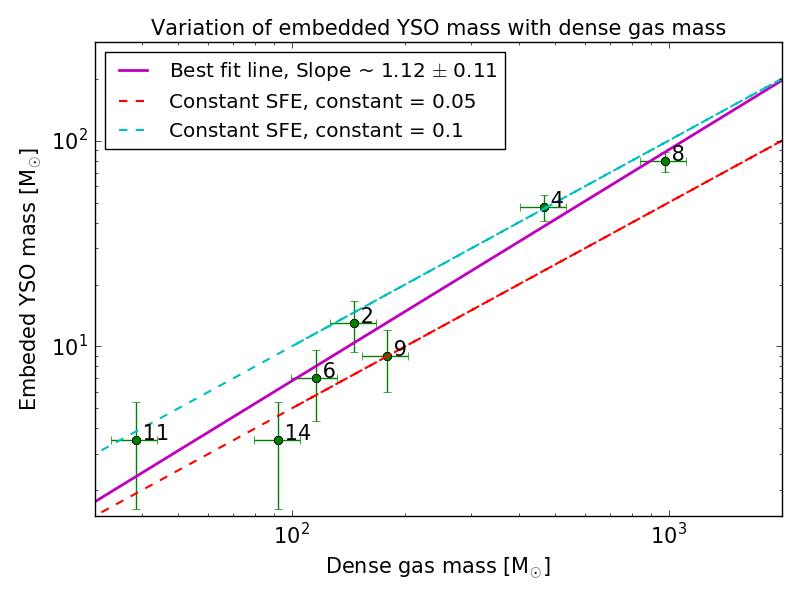}
\caption{Mass of embedded YSOs vs. dense gas mass for different regions.  Labels show the region numbers for each point, and regions that have no gas above 100 M$_{\odot} \rm{pc}^{-2}$ (and thus no "dense gas" nor "embedded YSOs") have been omitted.  The overplotted dash lines are the lines of constant star formation efficiency, and the solid line is the best fit line to the data.
\label{fig:ysoVsdg}}
\end{figure}

We also find a clear correlation between YSO mass and the total gas mass when we integrate over the N-PDF within the high column density regime. In figure 15, we plot the dense gas mass of each region versus its embedded YSO mass, tallying the gas and the number of YSOs within the $N(H_2) > 10^{22}$~cm$^{-2}$ contour (e.g. \citealt{2014ApJ...780..173B}). The YSO mass is calculated by assuming that each YSO is 0.5 M$_\odot$. The resulting integrated star formation efficiency (SFE) in this column density range is 5\% to 10\% as we progress from low dense gas to high dense gas masses. However, the requirement that the YSOs are observed coincident with the dense gas means that 30-80$\%$ of all YSOs are ignored in a given region. The slope of the best fit line in figure \ref{fig:ysoVsdg} is $\sim$1.12 $\pm$ 0.11, and the correlation is quite strong with a Pearson coefficient of 0.96. A constant dense gas SFE would give a slope near unity (e.g. \citealt{2010ApJ...724..687L}), while a SFE that rises with gas column density when integrated over the box would give a slope in excess of 1 \citep{2011ApJ...739...84G, 2013ApJ...778..133L}. A SFE that rises with gas column density would result from a star formation law where the star formation per area, $\Sigma_{star}$, increases as a power (index $>$ 1) of the gas surface density, $\Sigma_{gas}$. \cite{2011ApJ...739...84G} found that $\Sigma_{star} \propto \Sigma_{gas}^2$ in eight molecular clouds using the same Spitzer YSO data shown in this paper and using extinction maps derived for 2MASS to measure the gas column density. If we apply this law to the column densities shown for figure \ref{fig:fluxRatio} and then integrate to find the total mass of stars in each region, we find that this value gives a higher SFE ($\sim$ 15\% to 30\%) than that found for Mon R2 in figure \ref{fig:ysoVsdg}. However, we can reproduce the SFE between dense gas mass and embedded YSO mass in figure \ref{fig:ysoVsdg} ($\sim$ 5\% to 10\%) if we adopt the \cite{2011ApJ...739...84G} star-gas density correlation fit result for the Ophiuchus cloud where $\Sigma_{star} \propto \Sigma_{gas}^{1.8}$. There are two reasons why the Ophiuchus value may be more appropriate. First, the $2MASS$ near-IR data may have under predicted the gas column densities toward the densest regions of Mon R2, therefore biasing the power law exponent derived by \cite{2011ApJ...739...84G} power law index for Mon R2. Second, the physical beam size of the Herschel-derived column density map of Mon R2 is both much smaller than the corresponding beam size for the Mon R2 near-IR extinction map and similar in size to the beam size of the Ophiuchus near-IR extinction map.  From this analysis, we find the observed correlation between the mass of stars and mass of gas in Mon R2 is consistent with both a constant dense gas SFE and SFE that rises with column density, and uncertainties in these data are too large to distinguish these models.

Other interesting correlations are apparent. The regions with aggregates of filaments (i.e. those fit by the $G + 2$ N-PDFs) have large $x_{brk2}$ or shallow $\alpha_2$ ($>$ -3) and we see quite young stellar distributions relative to those with cluster-forming hubs ($G + 1$ N-PDFs). The Class~II/Class~I star count ratio is $\sim$3 for regions 6, 7, $\&$ 9 (those with $G + 2$ PDFs), in contrast to $\sim$5 for regions 2, 4 $\&$ 8 (those with $G + 1$). The Class II/Class I ratios of 3 are a signature of comparative youth, although they are not as low as that for the extremely young cluster Serpens South (Class II/Class I ratio of 0.77, \citealt{2008ApJ...673L.151G}). This result suggests either a longer protostellar lifetime or a later time since the onset of star formation in filament aggregate regions relative to regions containing cluster-forming hubs in Mon~R2. It remains an open question whether some filaments in a given aggregate could consolidate into cluster-forming hubs at a later epoch.

\begin{sidewaystable}[htbp]\centering
\setlength\tabcolsep{4.5pt}
\caption{best fitted parameter values corresponding to different accepted models for all 15 regions, as obtained using MCMC method. }
\label{tab:param}
  \begin{tabular}{cccccccccc}
  	\hline
     & Center (RA; Dec) &  &  & Log x$_0$  &  &  & Log $x_{brk1}$ &  & Log $x_{brk2}$ \\ 
    Region & [hh:mm:ss; hh:$'$:$''$] & Model & Log A & [cm$^{-2}$] & Log $\rm{\sigma}$ & $\alpha_1$ & [cm$^{-2}$] & $\alpha_2$ & [cm$^{-2}$] \\
    \hline
   
    1	& 6:14:41.530; -6:20:53.64 & $G+1$	& 3.272 $\pm$ 0.037	& 21.133 $\pm$ 0.017 	& 0.136 $\pm$ 0.007 	& -2.575 $\pm$ 0.018 	& 21.032 $\pm$ 0.002 	& ---	& --- \\
    
    2	& 6:12:45.773; -6:11:43.99 & $G+1$	&   3.105 $\pm$ 0.005 &  21.078 $\pm$ 0.002 &   0.104 $\pm$ 0.002 & -1.609 $\pm$ 0.015 &   21.190 $\pm$ 0.006   	& ---	& --- \\
    
	3	& 6:11:22.797; -5:43:43.26 & $G+2$	&   2.926 $\pm$ 0.005 &  21.007 $\pm$ 0.001  &   0.065 $\pm$ 0.001  & -4.408 $\pm$ 0.094  &  21.049 $\pm$ 0.007  &  -9.092 $\pm$ 0.399 &  21.238 $\pm$ 0.003 \\
	
	4	& 6:10:48.660; -6:11:46.46 & $G+1$	&   3.205 $\pm$ 0.006 &  21.049 $\pm$ 0.002  &   0.058 $\pm$ 0.003 & -1.476 $\pm$ 0.008  &  21.061 $\pm$ 0.004 	& ---	& --- \\
	
	5	& 6:11:10.362; -7:11:03.16 & $G+2$	&	2.873 $\pm$ 0.005   &  21.011 $\pm$ 0.001 &   0.087 $\pm$ 0.001  &  -0.678 $\pm$ 0.582 &  21.205 $\pm$ 0.002  &  -12.414 $\pm$ 1.164  &   21.282 $\pm$ 0.005 \\
	
	6	& 6:07:37.387; -5:10:43.81 & $G+2$	&	3.102 $\pm$ 0.029 &   20.963 $\pm$ 0.005  &   0.074 $\pm$ 0.004 &  -1.112 $\pm$ 0.008  & 20.911 $\pm$ 0.006  & -8.323 $\pm$ 1.099 & 22.086 $\pm$ 0.041 \\
	
	7	& 6:08:37.250; -5:54:02.27 & $G+2$	&	3.007 $\pm$ 0.003 &  21.279 $\pm$ 0.002  &   0.126 $\pm$ 0.003  &  -3.645 $\pm$ 0.243 &  21.287 $\pm$ 0.076 & -6.739 $\pm$ 0.642  & 21.588 $\pm$ 0.008 \\
	
	8	& 6:07:45.800; -6:21:43.55 & $G+1$	&   3.048 $\pm$ 0.003 &  21.409 $\pm$ 0.008  &  0.167 $\pm$ 0.007  &  -1.486 $\pm$ 0.038  &  21.627 $\pm$ 0.008	& ---	& --- \\
	
	9	& 6:08:04.324; -6:59:00.28 & $G+2$	&   3.051 $\pm$ 0.034  &   21.245 $\pm$ 0.007  &  0.105 $\pm$ 0.005  &  -0.701 $\pm$ 0.042  & 21.171 $\pm$ 0.019  & -2.518 $\pm$ 0.083 & 21.637 $\pm$ 0.005 \\
	
	10	& 6:09:00.134; -7:32:00.99 & $G+1$ &  2.919 $\pm$ 0.005 &   21.243 $\pm$ 0.003  &   0.131 $\pm$ 0.003  &  -3.965 $\pm$ 0.093  &  21.401 $\pm$ 0.018  &  ---  & --- \\
	
	11	& 6:05:54.363; -6:19:55.49 & $G+2$	&	3.105$\pm$ 0.084  &   21.142 $\pm$ 0.069  &  0.071$\pm$ 0.006  &  -1.621 $\pm$ 0.054 &   21.220 $\pm$ 0.004 	&  -2.37 $\pm$ 0.003	&  21.535 $\pm$ 0.056 \\
	
	12	& 6:06:24.493; -5:52:14.65 & $G+2$	&   3.231 $\pm$ 0.004  &  21.231 $\pm$ 0.003 &  0.076 $\pm$ 0.003 &  -1.849 $\pm$ 0.088 & 21.162 $\pm$ 0.004 & -3.142 $\pm$ 0.064  &  21.468 $\pm$ 0.002 \\
	
	13	& 6:05:54.207; -6:47:36.88 & $G+2$	& 2.894 $\pm$ 0.005 & 21.142 $\pm$ 0.002 & 0.078 $\pm$ 0.004	& -3.929 $\pm$ 0.006 & 21.197 $\pm$ 0.013	& -7.223 $\pm$ 0.008 & 21.371 $\pm$ 0.015 \\
	
	14	& 6:05:42.597; -7:16:00.39 & $G+2$	&   3.108 $\pm$ 0.005 &  21.135 $\pm$ 0.005  & 0.118 $\pm$ 0.004 &  -2.894 $\pm$ 0.025  &   21.227 $\pm$ 0.037	& -0.901 $\pm$ 0.005	& 21.667 $\pm $ 0.010\\
	
	15	& 6:03:42.738; -6:41:09.54 & $G+1$	&  2.952 $\pm$ 0.004  &  21.077 $\pm$ 0.001  &  0.120 $\pm$ 0.001 & -2.459 $\pm$ 0.158  &  21.359 $\pm$ 0.005	 & ---	& --- \\
	
	\hline
  \end{tabular}
\end{sidewaystable}

\begin{sidewaystable}[htbp]\centering
\setlength\tabcolsep{4.5pt}
\caption{Star-gas contents in each region.}
\label{tab:starGasProp}
  \begin{tabular}{ccccccccccc}
    \hline
    	&	&  & 	& YSO ratio	& ClassI YSOs	& ClassII YSOs	&  Total YSOs	&	Dense gas mass	&	Total gas mass 	& Dense gas mass fraction	\\
    Region & ClassI YSOs & ClassII YSOs & 	Total YSOs & ClassII/ClassI & Embedded & Embedded & Embedded & [M$_{\odot}$] & [M$_{\odot}$] & Dense gas / Total gas \\
    \hline
1	&	1	&	46	&	47	&	46	&	0	&	1	&	1	&	9.71	&	958.84	&	0.01\\
2	&	12	&	63	&	75	&	5.25	&	9	&	17	&	26	&	146.7	&	1384.96	&	0.11\\
3$^a$	&	0	&	3	&	3	&	$\infty$	&	0	&	0	&	0	&	0	&	748.7	&	0\\
4	&	23	&	127	&	150	&	5.52	&	20	&	75	&	95	&	467.95	&	1773.22	&	0.26\\
5$^a$	&	0	&	0	&	0	&	$\infty$	&	0	&	0	&	0	&	0	&	719.86	&	0\\
6	&	11	&	36	&	47	&	3.27	&	4	&	10	&	14	&	115.75	&	1392.83	&	0.08\\
7	&	4	&	12	&	16	&	3	&	0	&	0	&	0	&	0	&	1308.05	&	0\\
8	&	53	&	304	&	357	&	5.74	&	45	&	114	&	159	&	978.74	&	2998.21	&	0.33\\
9	&	19	&	46	&	65	&	2.42	&	9	&	9	&	18	&	178.98	&	2009.66	&	0.09\\
10$^a$	&	1	&	0	&	1	&	0	&	0	&	0	&	0	&	0	&	1300.25	&	0\\
11	&	7	&	36	&	43	&	5.14	&	4	&	3	&	7	&	38.6	&	1392.27	&	0.03\\
12	&	3	&	14	&	17	&	4.67	&	0	&	0	&	0	&	2.61	&	1433.23	&	0\\
13	&	1	&	4	&	5	&	4	&	0	&	0	&	0	&	0	&	1027.29	&	0\\
14	&	8	&	7	&	15	&	0.88	&	6	&	1	&	7	&	92.28	&	1276.31	&	0.07\\
15$^a$	&	2	&	3	&	5	&	1.5	&	0	&	0	&	0	&	0	&	870.22	&	0\\

	\hline
	\multicolumn{3}{l}{$^a$ Incomplete $Spitzer$ coverage}\\
  \end{tabular}
\end{sidewaystable}

\section{Conclusions and future work} \label{conclusion}

We present an analysis of the dust emission in the Mon~R2 GMC using $Herschel$. Data is reduced using $HIPE$ version 11.0.1 using $Planck-HFI$ for calibrating $SPIRE$ images. We performed a single temperature greybody fit to the 250 $\micron$, 350 $\micron$ and 500 $\micron$ $SPIRE$ data. The emissivity ($\beta$) can be used as a free parameter for fit but generally $\beta$ and temperature are degenerate unless we include higher wavelength data ($\sim$mm range). For this reason, generally in analyses like this $\beta$ is fixed beforehand and fits are performed for the remaining parameters. We used the flux ratio comparison plot to constrain a proper value of $\beta$ as a representative value for the whole cloud. We found this value to be 1.8. The flux ratio plot also gives a possibility of the $\beta$ values being scattered up to $\sim$1.5 on the lower limit and up to $\sim$ 2.0 on the upper limit. To constrain the systematic uncertainties of adopting a fixed $\beta$, we studied the variation in derived column densities and temperatures for taking these two extreme values of $\beta$.  We found that the shift would be $\sim$ 30$\%$, consistent with other dust continuum studies. We used a NIR extinction map \citep{2011ApJ...739...84G} to check the gas column density values that we derive from the dust emission maps, finding median gas density ratios of 0.95 and thus good agreement at intermediate column densities. We presented an analysis of the statistical confidence that our SED fits are distinct from the R-J limit, finding that those pixels with T $>$ 28K have a non-negligible probability of being on the R-J tail of the SED.  These pixels are discarded from our N-PDFs.
 
We studied the PDF of column densities in the whole cloud and found that the distribution is a lognormal for regions with column density $<$ 10$^{21}$ cm$^{-2}$ and changes to a power law form with slope -2.15 otherwise. Theoretically, supersonic turbulence is the responsible mechanism dominant over large scales in mostly diffuse low column density regions. This implies that the lognormal nature that we see for low column densities may be a consequence of supersonic turbulence. Similarly, on smaller scales the gas self-gravity wins over turbulence giving rise to a power law tail as seen in simulations. Hence, below the critical limit of $\sim$10$^{21}$ cm$^{-2}$ in Mon~R2 cloud, the regions may be better explained as turbulence-dominated regions and above this value the power law tail may be the consequence of the prevalence of self-gravity.

We extended our study further by selecting 15 sub-regions for N-PDF chatacerization to see if we observe similar behaviour in regional N-PDFs as in the whole Mon~R2 GMC. The regions were fixed in size and selected to span a range of YSO densitiy environments.  For all regions that contain dense YSO cluster cores, the N-PDF is a combination of a lognormal and one power law function. We do not see the power law excess in high column density region of the Mon~R2 cluster core, as reported by \cite{2015arXiv150708869S}. The regions with moderate numbers of YSOs and dense filamentary structures are better explained by a lognormal with two power laws. 

We studied how the power law index of N-PDFs, often claimed to be a defining factor of star formation, varies with the YSO count in several regions of Mon~R2. For the regions defined by a single power law, we found that the correlation is steeper than for the regions defined by two power laws.  In the latter case, the absence of high column density gas signifies lower star formation efficiency and thus less YSOs. While doing the regional analysis, we estimated the dense gas mass in each region and did a qualitative study of their relation with the YSO count that are embedded in the dense gas. We see a clear correlation of the dense gas mass with the embedded YSO count, but we lacked sufficient statistical constraint to differentiate between models that suggest the SFE is constant or rising with higher dense gas mass.

The emergence of the single power law in N-PDFs is often related with high star formation in those regions, but the presence of the second power law is still not fully understood. However, looking at the gas geometry in those regions, it seems to be related to the presence of dense filament aggregates. This leaves us with further open questions. Are these filament aggregates, which are represented in N-PDF by a shallow power law with steep cut-off at intermediate density, distinctly different structures from seemingly more monolithic clustered star forming sites that exhibit single power law N-PDFs?  Do the filament aggregates coalesce into the more monolithic cluster-forming sites, thereby evolving to reach a geometry where higher gas densities are achieved and stars can be formed considerably more efficiently?  A more detailed work that incorporates other nearby molecular clouds is required to address these questions and is beyond the scope of this paper. 



This work is based on observations made with $Herschel$, a European Space Agency cornerstone mission with significant participation by NASA. Support for this work was provided by NASA through an award issued by JPL/Caltech (contract number 1489384). S.~J. Wolk was supported by NASA contract NAS8-03060. We are thankful to Stella Offner, Mark Heyer, Grant Wilson and Ronald Snell from the University of Massachusetts (UMASS), Amherst for helpful conversations, suggestions, and feedback. We also thank Amy Stutz of the Max-Planck Institute for Astronomy, Germany for important suggestions on the paper. We also thank Bernhard Schulz and David Shupe from NASA $Herschel$ Science Center for helping us with data reduction. We are grateful to Manikarajamuthaly Sri Saravana and Joan Chamberlin for helping us with technical aspects. Finally, we would like to thank the anonymous referee for valuable comments and suggestions. $SPIRE$ has been developed by a consortium of institutes led by Cardiff Univ. (UK) and including Univ. Lethbridge (Canada); NAOC (China); CEA, LAM (France); IFSI, Univ. Padua (Italy); IAC (Spain); Stockholm Observatory (Sweden); Imperial College London, RAL, UCL-MSSL, UKATC, Univ. Sussex (UK); Caltech, JPL, NHSC, Univ. Colorado (USA). This development has been supported by national funding agencies: CSA (Canada); NAOC (China); CEA, CNES, CNRS (France); ASI (Italy); MCINN (Spain); SNSB (Sweden); STFC (UK); and NASA (USA). $PACS$ has been developed by a consortium of institutes led by MPE (Germany) and including UVIE (Austria); KUL, CSL, IMEC (Belgium); CEA, OAMP (France); MPIA (Germany); IFSI, OAP/AOT, OAA/CAISMI, LENS, SISSA (Italy); IAC (Spain). This development has been supported by the funding agencies BMVIT (Austria), ESA-PRODEX (Belgium), CEA/CNES (France), DLR (Germany), ASI (Italy), and CICT/MCT (Spain).

\bibliography{my_bib}

\bsp

\label{lastpage}

\end{document}